\documentclass[reqno]{amsart}
\usepackage{oldlfont,amssymb,appendix}
\newtheorem{lemma}{Lemma}
\newtheorem{satz}{Proposition}
\newtheorem{theorem}{Theorem}
\newtheorem{corolla}{Corollary}
\theoremstyle{definition}

\newtheorem{defi}{Definition}

\theoremstyle{remark}

\newtheorem{remark}{Remark}

\renewcommand{\labelenumi}{(\arabic{enumi})}
\newcommand{\tr}{\mathop{\mathrm{tr}}}

\DeclareMathOperator*{\ccon}{\overline{co}}
\begin{document}
\title[Playing with Fidelities]{Playing with Fidelities}
\author[P.M.\,Alberti]{Peter M. Alberti}
\address{Institute of Theoretical Physics\\
University of Leipzig\\
Augustusplatz 10, D-04109 Leipzig, Germany}
\email{Peter.Alberti@itp.uni-leipzig.de (peter.alberti@web.de)}
\begin{abstract}
The notion of partial fidelities as invented recently by A.\,Uhlmann
for pairs of finite dimensional density matrices will be extended to
the $vN$-algebraic context and is considered and thoroughly
discussed in detail from a mathematical point of view.
Especially, in the case of semifinite $vN$-algebras formulae and estimates for the partial
fidelity between the functionals of a dense cone of inner derived
normal positive linear forms are obtained. Also, some generalities on the notion  
of fidelity in quantum physics are collected in an appendix, and another system of 
mathematical axioms for fidelity over density operators, which is based on the concept     
of relative majorization and which is intimately related to complete positivity, is proposed.
\end{abstract}
\maketitle
\subsection*{Basic notions and settings}
In a $vN$-algebra $M$ over a Hilbert space ${\mathcal H}$ call nonzero
$x\in M_+$ {\em locally invertible} if the linear operator $y=x|_{s(x){\mathcal H}}$
is invertible within $s(x)Ms(x)$, where $s(x)$ is the support of $x$.
Let $s(x)^\perp={\mathbf 1}-s(x)$, with the unit operator ${\mathbf 1}$.
The {\em local inverse} $x^{-1}\in M_+$ of the locally invertible $x$ then is defined
as the unique element of $M$ obeying $x^{-1}|_{s(x){\mathcal H}}=y^{-1}$ and
$x^{-1}|_{s(x)^\perp{\mathcal H}}={\mathbf 0}$.

For simplicity, throughout the following we will tacitly agree
in considering only $vN$-algebras acting on {\em separable} Hilbert spaces.

In slightly modifying the settings from \cite{Uhlm:99.1,Uhlm:99.2}
in the finite dimensional case, we let a set $PAIRS(M)$ be defined
as set of all pairs $(a,b)$ of positive, locally invertible
elements of $M$ such that $aba=a$ and $bab=b$ hold. In addition, 
if $q\in M\backslash \{{\mathbf 0}\}$ is an orthoprojection,
we let $PAIRS_q(M)\subset PAIRS(M)$ be defined as
\begin{equation}\label{fundef}
PAIRS_q(M)=
\bigl\{(a,b)\in PAIRS(M):\ s(a)\approx s(b)\approx q\bigr\}\,.
\end{equation}
Thereby, `$\approx$' means `unitary equivalence' within $M$,
that is, for $x,y\in M$, $x\approx y$ holds if $y=u^*xu$ is fulfilled, for some
{\em unitary} $u\in M$.
Recall in this context the comparison relation `$\succ$' and the equivalence relation `$\sim$' for
orthoprojections of $M$.  For orthoprojections $p,q\in M$, $p\succ q$ (resp.~$p\sim q$) is
equivalent to $p\geq w^*w$ (resp.~$p=w^*w$) and $ww^*=q$, for some $w\in M$ (i.e.~$w$ is a
{\em partial} isometry). Thus especially, $p\sim q$ is equivalent to both $p\succ q$ and $q\succ p$.

For instance,
for $x\in M$ the left support $l(x)$ and right support $r(x)$
obey $l(x)\sim r(x)$. Also recall the notion of the modulus $|x|$ of an operator
$x\in M$, which is defined as $|x|=\sqrt{x^*x}$, and the
polar decomposition theorem for $x$, which states
that there exists a decomposition $x=w|x|$ with a partial isometry $w\in M$,
which is uniquely determined by the condition $w^*w=r(x)=s(|x|)$, and in which case then
$w w^*=l(x)=s(|x^*|)$ holds.

It is plain to see that over orthoprojections `$\approx$' in terms of
`$\sim$' equivalently can be also read as follows:\
$p\approx q$ holds if, and only if, both $p\sim q$ and $p^\perp\sim q^\perp$ are fulfilled.
Remind the reader to the (defining) pecularity through which a $vN$-algebra $M$ becomes a {\em finite}
$vN$-algebra: $M$ is finite if, and only if, the relation
$q\sim {\mathbf 1}$, with orthoprojection $q\in M$, has the unique solution  $q={\mathbf 1}$.
Together
with the comparability theorem for orthoprojections \cite[2.1.3.]{Saka:71} this fact implies that
for mutually equivalent orthoprojections $p\sim q$ in a finite $vN$-algebra also
$p^\perp\sim r^\perp$ holds, see \cite[2.4.2.]{Saka:71}. Hence,
in a finite $vN$-algebra equivalence $p\sim q$ of orthoprojections becomes equivalent to {\em unitary equivalence}
$p\approx q$, that is, $u^*q u=p$ and $up u^*=q$ then can be even achieved
with {\em unitary} operator
$u\in M$.

We start with an auxiliary result about stability of
local invertibility, essentially.
\begin{lemma}\label{locinv}
For each invertible $y\in M$ the following properties hold\textup{:}
\begin{enumerate}
\item\label{locinv1}
$s(y^*x y)\approx s(x),\,\forall\,x\in M_+$;
\item\label{locinv2}
if $x\in M_+$ is locally invertible, then $y^*x y$ is so, too;
\item\label{pairs.0}
$(a,b)\in PAIRS_q(M)\,\Longleftrightarrow\,(y^*ay,y^{-1}by^{{-1}*})\in PAIRS_q(M)\,.$
\end{enumerate}
\end{lemma}
\begin{proof}
In order to see \eqref{locinv1}, note first that $s(y^*x y)\sim s(\sqrt{x}y y^*\sqrt{x})
\leq s(x)$ is fulfilled, that is,
we have $s(x)\succ s(y^*x y)$. Also, in view of $s(x+s(x)^\perp)={\mathbf 1}$ and by invertibility
of $y^*$ one has
$\overline{\{x+s(x)^\perp\}^{1/2}{\mathcal H}}={\mathcal H}$ and
$y^*{\mathcal H}={\mathcal H}$.
Hence $\overline{y^*\{x+s(x)^\perp\}^{1/2}{\mathcal H}}={\mathcal H}$, and therefore 
$l(y^*\sqrt{x+s(x)^\perp})={\mathbf 1}$. Thus also
$s(y^*\{x+s(x)^\perp\} y)={\mathbf 1}$ must hold. Hence
$s(y^* s(x)^\perp y)\vee s(y^*x y)={\mathbf 1}$. We then may apply the parallelogram law for
orthoprojections \cite[2.1.5.]{Saka:71}, which is telling us that for orthoprojections $p,\,r$
in a $vN$-algebra
always $p\vee r - r\sim p- p\wedge r$ holds, with $p=s(y^* s(x)^\perp y)$
and $r=s(y^*x y)$ as follows\,:
$
{\mathbf 1}-s(y^*x y)\sim s(y^* s(x)^\perp y)-s(y^*x y)\wedge s(y^* s(x)^\perp y)
\leq s(y^* s(x)^\perp y)$.
Thus, for each $x\in M_+$ and any
invertible $y\in M$, we have $s(x)\succ s(y^*x y)$ and $s(y^* s(x)^\perp y)\succ s(y^*x y)^\perp$.
Thus, the former of the two relations especially must be fulfilled with the
special positive operator $s(x)^\perp$ instead
of $x$, and then reads as $s(x)^\perp\succ  s(y^* s(x)^\perp y)$.
By transitivity, the latter together with $s(y^* s(x)^\perp y)\succ s(y^*x y)^\perp$ then yields
$s(x)^\perp \succ s(y^*x y)^\perp$.
Hence, for any $x\in M_+$ and invertible $y\in M$ we
have $s(x)\succ s(y^*x y)$ and $s(x)^\perp \succ s(y^*x y)^\perp$.
Upon transforming simultaneously $x\,\mapsto\, y^*x y$ and
$y \,\mapsto\, y^{-1}$ the previous relations transform into $s(y^*x y)\succ s(x)$
and $s(y^*x y)^\perp \succ s(x)^\perp$. The latter together with the former then shows
that the equivalences
$s(x)\sim s(y^*x y)$ and $s(x)^\perp \sim s(y^*x y)^\perp$ must hold,
for each $x\in M_+$ and any invertible $y\in M$. This is the same as
$s(x)\approx s(y^*x y)$, and therefore \eqref{locinv1} is seen.

To see \eqref{locinv2}, assume $x\geq {\mathbf 0}$ is locally invertible, and denote $z=y^* xy$.
Note that without loss of generality we may
assume $x\leq {\mathbf 1}$. Local invertibility of such $x$ then is equivalent to existence of
$\varepsilon >0$ such that $s(x)\geq x\geq \varepsilon s(x)$. Hence, $y^*s(x) y\geq z\geq \varepsilon y^*s(x) y$,
that is, $s(z)=s(y^*s(x) y)$ must hold.
Let $s(x)y = v |s(x)y|$ be the polar decomposition of $s(x)y$. One then has $s(y^*s(x) y)=s(|s(x)y|)=r(s(x)y)=v^*v$
and $s(s(x)yy^*s(x))=s(|y s(x)|)=l(s(x)y)=vv^*$. From this $v y^*s(x) y v^*=s(x)yy^*s(x)$ can be obtained.

Now, invertibility of $y$ makes that $yy^*$ is positive invertible.
Thus there is $\delta >0$ with $yy^*\geq \delta {\mathbf 1}$. Hence $s(x)yy^*s(x)\geq \delta s(x)$, from which
in view of the previous $v y^*s(x) y v^* \geq \delta s(x)$ and $s(s(x)yy^*s(x))=s(x)$ can be seen.
From this in view of the above relations $ y^*s(x) y=v^*v y^*s(x) y v^* v\geq \delta v^*s(x) v=
\delta v^*s(|y s(x)|) v=\delta v^*v=\delta s(y^*s(x) y)=\delta s(z)$ is obtained. Thus we have
$ y^*s(x) y\geq \delta s(z)$. The latter and the above mentioned estimate
$ z\geq \varepsilon y^*s(x) y$
fit together into another estimate which is telling us that $z\geq \varepsilon\delta
s(z)$ has to hold. Owing to
$ \varepsilon\delta>0$ and
$\|z\| s(z)\geq z$ this then implies $z$ to be locally invertible, with $s(z)= s(y^*s(x) y)$.

Finally, relating \eqref{pairs.0}, note first that for invertible
$y\in M$ the map $\Phi_y$ defined within
$M\times M$ at the pair $(a,b)$ by $\Phi_y((a,b))=(y^*ay,y^{-1}by^{{-1}*})$ is
invertible, with $\Phi_y^{-1}=\Phi_{y^{-1}}$. Hence, for a complete proof of
\eqref{pairs.0} one may content with showing that the
`$\Longrightarrow$'-\,direction is valid, for each invertible $y\in M$. This
we are going to do now. Let $(a,b)\in PAIRS_q(M)$. In view of the definition of $PAIRS_q(M)$,
by applying \eqref{locinv1} with invertible $y$ and $(y^{-1})^*$ to $x=a$ and $x=b$, respectively,
we see that $s(a)\approx s(y^*a y)$ and
$s(b)\approx s(y^{-1} b {y^{-1}}^*)$ have to be fulfilled. Hence, $s(a)\approx s(b)\approx q$ also implies $s(y^*a y)\approx s(y^{-1} b {y^{-1}}^*)\approx q$.
Also, by \eqref{locinv2} both $y^*a y$ and
$ y^{-1} b{y^{-1}}^*$ are locally invertible (since $a$ and $b$ are so). And finally, since
$(a,b)\in PAIRS(M)$ also means that $aba=a$ and $bab=b$ hold, the latter kind of relations
owing to $(y^*a y)(y^{-1} b{y^{-1}}^*)(y^*a y)=y^*aba y=y^*a y$ and
$(y^{-1} b{y^{-1}}^*)(y^*a y)(y^{-1} b{y^{-1}}^*)=y^{-1} bab{y^{-1}}^*=y^{-1} b{y^{-1}}^*$
is reproduced to hold also between $y^*a y$ and $y^{-1} b{y^{-1}}^*$. Thus
$(y^*a y,y^{-1} b{y^{-1}}^*)\in  PAIRS_q(M)$.
\end{proof}
Suppose $\omega,\varrho\in M_+^*$ to be positive linear forms on the $vN$-algebra $M$,
and be $q$ an orthoprojection. In affinity with Uhlmann's proposal for $n\times n$-matrices, in the $vN$-algebraic
environment define the
{\em $q$-partial fidelity} of the pair $\{\omega,\varrho\}$ as follows:
\begin{defi}[after {\sc A.\,Uhlmann}, \cite{Uhlm:99.1}]\label{pafi}
$$F_M(\omega,\varrho|q)=\left\{
\begin{array}{ll} \inf_{(a,b)\in PAIRS_q(M)}\frac{1}{2}\{\omega(a)+\varrho(b)\} & \text{ for }q\not={\mathbf 0}, \\
& \\
0 & \text{ for } q={\mathbf 0}\,.\end{array}\right.$$
\end{defi}
Subsequently, each functional of the family
$\bigl\{F_M(\cdot,\cdot|q):q\in M,\text{orthoprojection}\bigr\}$ will be referred to as
{\em partial fidelity}, simply. By definition, partial fidelities can be labelled by the
classes of unitary equivalent orthoprojections within $M$. If $q={\mathbf 1}$ happens, we refer to
$F_M(\omega,\varrho|{\mathbf 1})$ as {\em fidelity} of the pair $\{\omega,\varrho\}$, in which case
then also the notation $F_M(\omega,\varrho)$ will be used.
\subsection*{Some properties of fidelity}
Start with some quite incomplete remarks on the notion of fidelity.
In communication theory the concept of
`fidelity' refers to some quantitative measure(s) $F$, with
$0\leq F(\varrho_{\text{in}},\varrho_{\text{out}})\leq 1$, and estimating the
accuracy of transmission through some (specific) communication channel. Thereby,
$\varrho_{\text{in}}$ and $\varrho_{\text{out}}$ refer to the input and output state of the
channel, accordingly. Clearly, which states have to be considered
to be close to each other essentially depends on the conceptual frame and
mathematical notion of state used therein, and `fidelity' then is intended to
evaluate quantitatively this topological notion of neighbourhood.

To be qualified as such a
measure it is reasonable to impose on $F$ some canonical
structural properties. There are at least
two aspects to be considered in that context: {\em functional} and
{\em functorial} ones. The latter reflect that `fidelity' should be
applicable to a whole category of (similar)
systems (channels), and thus one rather has to do with fidelity functors than with
fidelity functions, merely. Examples of more functorial properties are the above required
$0\leq F(\varrho_{\text{in}},\varrho_{\text{out}})\leq 1$ and, yet more important,
that $ F(\varrho_{\text{in}},\varrho_{\text{out}})=1$ should be equivalent to
$\varrho_{\text{in}}=\varrho_{\text{out}}$ (best accuracy of transmission).
On the other hand,  in order to meet the needs of
{\em quantum} communication
theory for instance (in which we are in a $vN$-algebraic context),
convexity and subadditivity properties (joint concavity, e.g.) might be considered as
reasonable features of fidelity $F$.

But note that in general neither a single fidelity functor $F$
will be uniquely determined by those properties nor will a best
(minimal) sufficient system of fidelity functors exist. For the
needs of {\em quantum} communication theory, R.~Jozsa in
\cite{Jozs:94} has proposed a system of axioms for fidelity. A
similar but slightly differing system of axioms is proposed and
discussed in the appendix to this paper. Interestingly, both
systems of axioms (and not only these) admit least functors of
fidelity, which both descend from the functor of the
$^*$-algebraic transition probability $P_M$ in the sense of
\cite{Uhlm:76}. Much about this exposed functor is known
\cite[(1.1)--(1.9)]{AlUh:84}. Thus, if we refer to {\em joint
concavity} as a desirable property (see Appendix), then the least
functor reads as $F_M=\sqrt{P_M}$. Note that $P_M$ can be defined
on all pairs of positive linear forms. From now on and for
simplicity, when speaking of `fidelity', the expression
$F_M=\sqrt{P_M}$ will be meant, with arguments extending through
$M_+^*$.

In this sense, from \cite[\sc{Corollary} 3\,(5)]{aluh:00.1} e.g.~one infers that
the following remarkable representation property holds\,:
\begin{subequations}\label{fi}
\begin{equation}\label{fi.1}
F_M(\omega,\varrho)=\inf_{a\in M_+,\,\text{invertible}}\frac{1}{2}\{\omega(a)+\varrho(a^{-1})\}\,.
\end{equation}
Since locally invertible elements of full support are the invertible positive elements, we
have $PAIRS_{\mathbf 1}(M)=\bigl\{(a,a^{-1}):a\in M_+,\text{invertible}\bigr\}$, and by
the previous formula and
in accordance with Definition \ref{pafi} we then in fact have
$F_M(\omega,\varrho|{\mathbf 1})=F_M(\omega,\varrho)=
\sqrt{P_M(\omega,\varrho)}$.

Start with some properties of fidelity which follow from the basic properties of $P_M$.
Thus one knows that the following
overall-estimate for the value of the fidelity by the functional
norm of the involved positive linear forms is fulfilled\,:
\begin{equation}\label{fi.1a}
F_M(\omega,\varrho)\leq \sqrt{\omega({\mathbf 1})\varrho({\mathbf 1})}\,,
\end{equation}
and if $a,b,c,d\in M$ obey $a^*b=c^*d$, then
\cite[\sc{Theorem} 1]{aluh:00.1} implies
\begin{equation}\label{fi.2}
F_M(\omega^a,\varrho^b)=F_M(\omega^c,\varrho^d)\,,
\end{equation}
where for $z\in M$ the linear form $\omega^z$ is defined as $\omega^z=\omega(z^*(\cdot)z)$,
and in which case
$\omega^z$ will be also referred to as {\em inner-derived} (from $\omega$) positive linear form.
There is also a special subadditivity property known in context of inner derived
positive linear form and fidelity. Namely, if $a^*b=\sum_{j=1}^n a_j^*b_j$ holds within $M$ ($n<\infty$),
then
\begin{equation}\label{fi.2'}
F_M(\omega^a,\varrho^b)\leq \sum_{j=1}^n F_M(\omega^{a_j},\varrho^{b_j})
\end{equation}
is fulfilled, see \cite[\sc{Corollary 3},\,{\em Remark 2}\,(3)]{aluh:00.1}.

It is useful to know yet two other basic properties of fidelity which follow
from the presupposed relation $F_M=\sqrt{P_M}$\,: {\em continuity} and
{\em hereditarity}. That is, whenever for positive linear forms one has
$\omega_n\to \omega$  and $\varrho_m\to\varrho$ uniformly, and if $q$ is an orthoprojection, then
\begin{equation}\label{fi.2a}
F_M(\omega,\varrho)=\lim_{n,m\to\infty} F_M(\omega_n,\varrho_m)
\end{equation}
holds, respectively the following fact is true:
\begin{equation}\label{fi.3}
F_M(\omega^q,\varrho^q)=F_{qMq}(\omega|_{qMq},\varrho|_{qMq})\,.
\end{equation}

In some special cases fidelities can be calculated easily. Thus, the main result of \cite{Uhlm:76}
when read in terms of fidelity tells us that for each positive linear form $\mu$ the following is fulfilled:
\begin{equation}\label{fi.5}
a,b\in M,\,a^*b\geq {\mathbf 0}\,:\ F_M(\mu^a,\mu^b)=\mu(a^*b)\,.
\end{equation}
Knowing this special case and \eqref{fi.2} one can even conclude that in case of $a,b\in M$ which
combine together to an element $a^*b$ of the so-called $\mu$-centralizer
$M^\mu$, which reads as
$M^\mu=\{y\in M:\mu(xy)=\mu(yx),\,\forall x\in M\}$, one has the following to hold\,:
\begin{lemma}\label{cent}
Let $\mu$ be a positive linear form, and $a,b\in M$ with $a^*b\in M^\mu$. Then
\begin{equation}\label{fi.6}
F_M(\mu^a,\mu^b)=\mu(|a^*b|)\,,
\end{equation}
is fulfilled.
\end{lemma}
\begin{proof}
First note that $M^\mu$ is always a unital ${\mathsf C}^*$-subalgebra of $M$. Hence, if $a^*b\in M^\mu$
one also has $b^*a=(a^*b)^*\in M^\mu$ and $|a^*b|\in M^\mu$. But then we may use \eqref{fi.2} and conclude as follows:
$F_M(\mu^a,\mu^b)=F_M(\mu,\mu^{a^*b})=F_M(\mu,\mu(|b^*a|^2(\cdot)))=F_M(\mu,\mu^{|b^*a|})$. Since with the help of
\eqref{fi.5} the relation $F_M(\mu,\mu^{|b^*a|})=\mu(|b^*a|)$ can be seen, $F_M(\mu^a,\mu^b)=\mu(|b^*a|)$ must
be valid.
Finally, by definition $F_M$ is a symmetric setting, that is, $F_M(\omega,\varrho)=F_M(\varrho,\omega)$.
These facts yield \eqref{fi.6}.
\end{proof}
Particularly interesting is the case if $\mu$ is a {\em tracial} positive linear form (finite trace),
$\mu=\tau$, that
is, $\tau\in M_+^*$ with $\tau(x^*x)=\tau(xx^*)$, for all $x\in M$ (property of {\em invariance}).
In this case
one has $M^\mu=M$, and thus formula \eqref{fi.6} then reads as
\begin{equation}\label{fi.6a}
\forall\,a,b\in M\,:\  F_M(\tau^a,\tau^b)=\tau(|a^*b|)\,.
\end{equation}
\end{subequations}
These formulae together with continuity \eqref{fi.2a} of fidelity, see e.g.~\cite[(2.4)]{AlUh:84} and
\cite[\sc{Proposition} 1]{aluh:00.1}, are quite useful in calculating fidelities also in other cases.
\subsection*{Extending Uhlmann's formula to the $vN$-case}
In the finite dimensional factor case, that is if $M$ is
isomorphic to the full algebra of $n\times n$-matrices, $M\simeq
{\mathsf M}_n({\mathbb C})$, A.\,Uhlmann \cite{Uhlm:99.1,
Uhlm:99.2} proved that, for two positive
semidefinite matrices $\omega$ and $\varrho$,
the sum of the $k$ smallest eigenvalues of
$|\sqrt{\omega}\sqrt{\varrho}\,|$ (if eigenvalues are counted
according to their multiplicity) exactly equals the partial
fidelity $F_M(\omega,\varrho|r)$ as defined in accordance with
Definition \ref{pafi}, provided $\omega,\,\varrho$ within the
latter definition are identified with the linear forms
$\omega(\cdot)=\tau^{\sqrt{\omega}}$ and
$\varrho(\cdot)=\tau^{\sqrt{\varrho}}$, where $\tau=\tr$ is the
standard trace on $n\times n$-matrices, and $r$ can be any
orthoprojection obeying ${\mathrm rank}(r)=\tr r = k$. On the
other hand, by equivalently reformulating a classical result of
matrix theory \cite{KyFa:51} one easily sees that, whenever
$$\mu_1(a)\geq \mu_2(a)\geq \,\ldots\,\geq \mu_n(a)$$ is the
ordered sequence of the singular values of an $n\times n$-matrix
$a$, for the sum of the $k$ smallest singular values the following
expression is obtained: $$\sum_{j>n-k} \mu_j(a)=\inf_{{\mathrm
rank}(q) = k,\,q \text{--orthoprojection}} \tr|a|q\,. $$ In
summarizing from these facts, what has been shown in \cite{Uhlm:99.1, Uhlm:99.2} amounts to
the fact that within
$n\times n$-matrices and for $r$ with ${\mathrm rank}(r)=k$ the
following holds:
\begin{equation*}
F_M(\omega,\varrho|r)=\inf_{{\mathrm
rank}(q) = k,\,q \text{--orthoprojection}}
\tr|\sqrt{\omega}\sqrt{\varrho}\,|q\,.
\end{equation*}
Since ${\mathrm rank}(q)={\mathrm rank}(r)$ means unitary equivalence, the previous formula equivalently also reads
$F_M(\omega,\varrho|r)=\inf_{q\approx r} \tr|\sqrt{\omega}\sqrt{\varrho}\,|q$.
What will be shown in this paper is that the latter formula  in
adaptation to a $vN$-algebraic context remains true.
\begin{theorem}\label{haupt}
Let $\tau$ be a normal trace on a $vN$-algebra $M$. For each
$x,y\in M$ and obeying $\tau(x^*x)<\infty$ and $\tau(y^*y)<\infty$ the following formula holds:
\begin{equation}\label{finfid.1}
F_M(\tau^x,\tau^y|r)=\inf_{q\approx r}  \tau(|x^*y|q)\,.
\end{equation}
Thereby, if $M$ is finite the infimum is a minimum.
\end{theorem}
\subsection*{Traces and fidelity}
The first step towards a proof of Uhlmann's formula
\eqref{finfid.1} will be to ask whether
formula \eqref{fi.6a} for fidelity remains true for {\em traces} on a $vN$-algebra.

Let $\tau$ be a trace on $M_+$. Then, ${\mathcal L}^2(M,\tau)=\{x\in M:\tau(x^*x)<\infty\}$ and
${\mathcal L}^1(M,\tau)=\{x\in M:\,x=y^*z,\ y,z\in {\mathcal L}^2(M,\tau)\}$ both are
$^*$-ideals with the same closure. Thereby, the ideal of all $\tau$-trace-class operators ${\mathcal L}^1(M,\tau)$
is the complex linear span of the hereditary positive cone
${\mathcal L}^1(M,\tau)_+=\{x\in M_+:\,\tau(x)<\infty\}$, and therefore $\tau$ uniquely
extends to a positive linear form $\tilde{\tau}$ on the $^*$-algebra ${\mathcal L}^1(M,\tau)$
(for basic facts see e.g.~in \cite[6.1.]{Dixm:64}), which is {\em invariant}. The latter means that
$\tilde{\tau}(xy)=\tilde{\tau}(yx)$ be fulfilled, for either $x,y\in {\mathcal L}^2(M,\tau)$, or $x\in M$ and
$y\in {\mathcal L}^1(M,\tau)$. Owing to uniqueness, $\tau$ often
will be identified with the linear form $\tilde{\tau}$, and then in this context of linear forms
is referred to as
{\em trace $\tau$ on $M$}. Having in mind these facts one at once infers that the previously introduced notion
$\tau^z=\tau(z^*(\cdot)z)$
of an inner derived positive linear form $\tau^z$ makes sense also with
a trace $\tau$ provided
$z\in {\mathcal L}^2(M,\tau)$ is fulfilled. In tacitely supposing the latter
the above notion will be made use of henceforth also in this extended sense.

Recall that a trace $\tau$ is {\em lower semi-continuous} whenever for each $x\in M_+$ and net
$\{x_\lambda\}\subset M_+$ obeying $x=\|\cdot\|-\lim_{\lambda} x_\lambda$ the relation
$\tau(x)\leq \liminf_\lambda \tau(x_\lambda)$ follows. Lower semi-continuity for a trace implies
$\tau(x)=\sup \{\tau(y):\,x\geq y\geq {\mathbf 0},\,\tau(y)<\infty\}$ to be fulfilled,
for each $x\in \overline{{\mathcal L}^1(M,\tau)}$, which is a weakened form of {\em semi-finiteness}
(usually in addition then $M=\overline{{\mathcal L}^1(M,\tau)}$ is assumed,
see \cite[6.1.3.]{Dixm:64} for details).

Remind that {\em finite} traces
as well as {\em normal} traces are examples of lower semi-continuous traces.
The announced modification of \eqref{fi.6a} then reads as follows\,:
\begin{corolla}\label{centcor}
Let $\tau$ be a lower semi-continuous trace, and $a,b\in {\mathcal L}^2(M,\tau)$. Then
\begin{equation}\label{fi.7}
F_M(\tau^a,\tau^b)=\tau(|a^*b|)
\end{equation}
is fulfilled. Moreover, if $s(|a|)\vee s(|b|)\in{\mathcal L}^1(M,\tau)$ is fulfilled, the formula
remains true also without assuming lower semi-continuity for $\tau$.
\end{corolla}
\begin{proof}
Let $I(M,\tau)=\{x\in M:\,\tau(x^*x)=0\}$. This is a $^*$-ideal
of $M$ which belongs to ${\mathcal L}^2(M,\tau)$. We now consider the
completion $L^2(M,\tau)$ of the linear space of all
equivalence classes $\xi_x$ to elements
$x\in {\mathcal L}^2(M,\tau)$ modulo $I(M,\tau)$ under the inner product
$\langle \xi_x, \xi_y\rangle=\tau(y^* x)$. A line of standard arguments ensures
that the latter is a well-defined setting and in fact extends to a scalar product
over $L^2(M,\tau)$ under which
this space gets a Hilbert space. We let $\pi: M\in x\,\longmapsto\,\pi(x)$ and
$\pi^{\prime}(x): M\in x\,\longmapsto\,\pi^{\prime}(x)$ be the
$^*$-representations of $M$ over $L^2(M,\tau)$
which can be uniquely given through $\pi(x)\xi_y=\xi_{xy}$ and
$\pi^{\prime}(x)\xi_y=\xi_{yx}$, respectively, for each
$y\in {\mathcal L}^2(M,\tau)$ and all $x\in M$.
Then, for each $z\in {\mathcal L}^2(M,\tau)$ and $x\in M$ one has
$\tau^z(x)=\langle \pi(x)\xi_{z},\xi_{z}\rangle$. Note that $\pi^{\prime}(x)\in
\pi(M)^{\,\prime}$ (commutante of $\pi(M)$). Thus, especially for each
partial isometry $w\in M$, one has that $\pi^{\prime}(w)\in
(\pi(M)^{\,\prime})_1$ holds, where $(\pi(M)^{\,\prime})_1$ is the unit sphere
of $\pi(M)^{\,\prime}$. Also, by definition of $P_M$ in \cite{Uhlm:76} and according to
another characterization of the transition probability in \cite{Albe:83}, for
each $w\in M$ with $\|w\|\leq 1$ one has $F_M(\tau^a,\tau^b)=\sqrt{P_M(\tau^a,\tau^b)}\geq
|\langle \xi_b,\pi^{\prime}(w)\xi_{a}\rangle |=|\langle \xi_b,\xi_{aw}\rangle |
=|\tau(w^* a^*b)|$.
Now, let $a^*b=v|a^*b|$ be the polar decomposition of $a^*b$. Then one has
$v^*v=s(|a^*b|)$. Thus the previous estimate may be considered for $w=v$, in
which case the following estimate arises\,:
\begin{subequations}\label{untob}
\begin{equation}\label{unter}
F_M(\tau^a,\tau^b)\geq \tau(|a^*b|)\,.
\end{equation}
Now, suppose $q$ is an orthoprojection with $0<\tau(q)<\infty$. Then, for $\tau^q\in M_+^*$ one has
$qMq\subset M^{\tau^q}$. Thus in particular $(aq)^*(bq)=q(a^*b)q\in M^{\tau^q}$, and therefore
Lemma \ref{cent} can be applied with $\mu=\tau^q$ and $aq,\,bq$ instead of $a,\,b$, respectively, and
then yields that
$F_M(\tau^{aq},\tau^{bq})=F_M((\tau^q)^{aq},(\tau^q)^{bq})=(\tau^q)(|(aq)^*(bq)|)=\tau(|q(a^*b)q|)$.
Now, by a simple monotony argument $q\leq {\mathbf 1}$ implies
$|q(a^*b)q|^2=(qb^*a)q(a^*bq)\leq q|a^*b|^2 q$. From this by operator monotony
of the square root (for generalities on that
see \cite{BeSh:55,Dono:74}, and \cite{Ando:78} for a nice survey) one infers $|q(a^*b)q|\leq \sqrt{q|a^*b|^2 q}=
|\,|a^*b|q|$. Note that invariance of $\tau$ especially also makes that $\tau(|z|)=\tau(|z^*|)$, for each
$z\in {\mathcal L}^1(M,\tau)$. Hence, for $z=|a^*b|q$ and by positivity of $\tau$
from the previous $\tau(|q(a^*b)q|)\leq \tau(|\,|a^*b|q|)=\tau(|q|a^*b|\,|)$ can be seen. Once more again
argueing by the same kind of monotony and operator monotony argument, we also see that
$|q|a^*b|\,|\leq |a^*b|$ holds, from which $\tau(|q|a^*b|\,|)\leq \tau(|a^*b|)$ follows. Taking together all these
estimates shows that the following has to be fulfilled\,:
\begin{equation}\label{oben}
\forall\,q\in {\mathcal L}^1(M,\tau),\text{ orthoprojection}\,:\ \tau(|a^*b|)\geq F_M(\tau^{aq},\tau^{bq})\,.
\end{equation}
\end{subequations}
Now, since according to $a,b\in {\mathcal L}^2(M,\tau)$ one has $a^*a + b^*b\in {\mathcal L}^1(M,\tau)_+$,
by some spectral calculus in $vN$-algebras one can be assured that there exists an ascendingly
directed sequence $\{q_n\}$ of orthoprojections, each of which commutes with $a^*a + b^*b$,
and which, for all $n\in {\mathbb N}$, obey
$\tau(q_n)<\infty$ and fulfil $\bigvee_n q_n=s(a^*a+b^*b)$ (lowest upper bound) and
$\|\cdot\|-\lim_{n\to\infty} q_n (a^*a+b^*b)=a^*a + b^*b$. On the other hand, by invariance and some
monotony and positivity argument
\begin{equation}\label{nconv}
\|\tau^a-\tau^{aq_n}\|_1+\|\tau^b-\tau^{bq_n}\|_1=\tau(q_n^\perp(a^*a+b^*b))
\tag{$\sharp$}
\end{equation}
has to hold. Note that by assumptions on $\{q_n\}$ and owing to monotony and positivity of $\tau$
from $q_n (a^*a+b^*b)=
\sqrt{a^*a+b^*b}\, q_n \sqrt{a^*a+b^*b}\leq a^*a+b^*b$ the relation $\tau(q_n (a^*a+b^*b))\leq
\tau(a^*a+b^*b)$ follows, for each $n\in {\mathbb N}$. From this in view of
$\|\cdot\|-\lim_{n\to\infty} q_n (a^*a+b^*b)=a^*a + b^*b$ and by lower semi-continuity of $\tau$ then
$\lim_{n\to\infty} \tau(q_n^\perp(a^*a+b^*b))=0$ follows.
Hence, by \eqref{nconv}
both $\tau^{aq_n}\,\to\,\tau^a$ and
$\tau^{bq_n}\,\to\,\tau^b$ uniformly, and in view of \eqref{fi.2a} therefore we arrive at
$\lim_{n\to\infty}  F_M(\tau^{aq_n},\tau^{bq_n})=F_M(\tau^a,\tau^b)$. In view of eqs.~\eqref{untob}, after
substitution of $q=q_n$ into \eqref{oben} accordingly and taking the
limit, the validity of \eqref{fi.7} then gets obvious.

Note that both estimates in \eqref{untob} were derived without making use of the assumption of
lower semi-continuity for $\tau$, and thus both have to hold with an arbitrary trace.
Under the stronger assumption of $s(|a|)\vee s(|b|)\in {\mathcal L}^1(M,\tau)$ owing to
$s(|a|)\vee s(|b|)\in {\mathcal L}^2(M,\tau)$
we therefore may choose $q=s(|a|)\vee s(|b|)$ in \eqref{oben}, and owing to
$F_M(\tau^{aq},\tau^{bq})=F_M(\tau^a,\tau^b)$ the conclusion from \eqref{untob} then again
provides equality, and thus yields formula \eqref{fi.7}, for an arbitrary trace.
\end{proof}
According to formula \eqref{fi.7}, fidelity $F_M(\tau^x,\tau^y)$ is the norm
of the positive linear form $\tau(|x^*y|(\cdot))$.  For normal traces, the latter
form subsequently will be shown to play a major r{\^o}le in identifying partial fidelity
$F_M(\tau^x,\tau^y|r)$. Thus, for later use and since it can be dealt with by
essentially the same means which just have been used in the previous proof,
add some consideration on continuity of this form.
\begin{lemma}\label{confi}
Let $\tau$ be a normal trace on $M$, $a,b\in  {\mathcal L}^2(M,\tau)$, and let
$\{q_n\}\subset M$ be an ascending sequence of orthoprojections which strongly
tends to unity. Then, with $a_n=aq_n$, for each $n\in {\mathbb N}$, one has
\begin{enumerate}
\item\label{confi1}
$\|\cdot\|_1-\lim_{n\to\infty}
\tau(\bigl|a_n^*\,b\bigr|(\cdot))=\tau(|a^*b|(\cdot))\,.$
\end{enumerate}
Provided $s(|a|)\in {\mathcal L}^1(M,\tau)$, for $b\geq {\mathbf 0}$
and with $b_n=|bq_n|$, for each $n\in {\mathbb N}$, one has
\begin{enumerate}
\setcounter{enumi}{1}
\item\label{confi2}
$\|\cdot\|_1-\lim_{n\to\infty}
\tau(\bigl|a^*\,b_n\bigr|(\cdot))=\tau(|a^*b|(\cdot))\,.$
\end{enumerate}
\end{lemma}
\begin{proof}
Note that for $z\in {\mathcal L}^1(M,\tau)$ the modul $|\tau(z(\cdot))|$ of the normal linear form
$\tau(z(\cdot))$ obeys $|\tau(z(\cdot))|=\tau(|z|(\cdot))$. Hence, according to
\cite[Prop.~4.10]{Take:79} the assertions of \eqref{confi1} and \eqref{confi2} will
follow if $\|\cdot\|_1-\lim_{n\to\infty}
\tau(a_n^*b(\cdot))=\tau(a^*b(\cdot))$ and $\|\cdot\|_1-\lim_{n\to\infty}
\tau(a^*b_n(\cdot))=\tau(a^*b(\cdot))$ can be shown to hold, under the respective premises
on $a,b,a_n$ or $a,b,b_n$, accordingly. This will be done now.

In case of \eqref{confi1}, note that the fact to be shown is equivalent to the relation
$\lim_{n\to\infty}\tau(|\{a-a_n\}^*b|)=0$. Now, by polar decomposition,
$|\{a-a_n\}^*b|=v_n^*\{a-a_n\}^*b=v_n^*q_n^\perp a^*b$, for some partial isometry $v_n$.
Hence, owing to $q_n^\perp\downarrow
{\mathbf 0}$ strongly, and since $\|\,|\{a-a_n\}^*b|\varphi\|=\|v_n^*q_n^\perp a^*b\varphi\|
\leq \|q_n^\perp a^*b\varphi\|$, for each $\varphi\in {\mathcal H}$, $|\{a-a_n\}^*b|\to {\mathbf 0}$ strongly is seen. Moreover, by operator monotonicity of the square root, from $|\{a-a_n\}^*b|=
\sqrt{b^*a q_n^\perp a^*b}$ even $|\{a-a_n\}^*b|\downarrow {\mathbf 0}$ strongly follows.
Thus $\lim_{n\to\infty}\tau(|\{a-a_n\}^*b|)=0$, by normality of $\tau$.

In case of \eqref{confi2}, the fact to be shown is
equivalent to $\lim_{n\to\infty}\tau(|a^*\{b_n-b\}|)=0$. In order to see the latter,
remark first that for $p_n=s(|a^*\{b_n-b\}|)$ we must have $p_n\prec s(|a|)$. Hence, by
assumption on $a$ we also have $\tau(p_n)<\infty$, and therefore by a common
Cauchy-Schwarz estimate
the following can be obtained:
$
\tau(|a^*\{b_n-b\}|)^2\leq \tau(p_n)\cdot \tau(|a^*\{b_n-b\}|^2)\leq \tau(s(a))\cdot\tau(aa^*\{b_n-b\}^2)
\leq  \tau(s(a))\|a\|^2\cdot\tau(\{b_n-b\}^2)$.
We are going to show $\limsup_{n\to\infty}\tau(\{b_n-b\}^2)=0$.
First note that by monotony of positive maps and operator monotony of the square root
from $q_n\leq {\mathbf 1}$
the estimate $b_n=\sqrt{q_nb^2q_n}\geq \sqrt{q_nbq_n b q_n}=q_nbq_n$ can be obtained. Thus,
by positivity of $\tau(b(\cdot))$ over ${\mathcal L}^2(M,\tau)$,
$\tau(bq_nbq_n)\leq \tau(b\,b_n)$ follows. But then, for each fixed $m\in {\mathbb N}$ and all
$n\geq m$,
$\tau(\{b_n-b\}^2)=\tau(b^2q_n)+\tau(b^2)-2\tau(b\,b_n)\leq \tau(b^2q_n)+\tau(b^2)-
2\tau(bq_nbq_n)\leq
\tau(b^2q_n)+\tau(b^2)-2\tau(bq_mbq_n)$. By normality of $\tau$, both $\tau(b^2(\cdot))$ and
$\tau(bq_mb(\cdot))$ are normal positive linear forms.
Thus, in view of $\bigvee_n q_n={\mathbf 1}$,
$
\limsup_{n\to\infty}\tau(\{b_n-b\}^2)\leq 2\tau(b^2q_m^\perp)
$
follows, for each $m\in {\mathbb N}$. Once more again by normality of $\tau(b^2(\cdot))$
and because of $q_m^\perp\downarrow {\mathbf 0}$ strongly, $\limsup_{n\to\infty}\tau(\{b_n-b\}^2)=0$
is seen.
\end{proof}

\subsection*{Some facts about $PAIRS_r(M)$}
In the following, a pair $\{p,q\}$ of orthoprojections $p,q\in M$ is referred to as a {\em minimal pair of
orthoprojections} if $s(pqp)=p$ and $s(qpq)=q$ hold and both $|qp|$ and $|pq|$ are locally invertible.
Let $qp=v(p,q)|qp|$ be the polar decomposition of $qp$, with
the partial isometry $v=v(p,q)$. For a minimal pair $\{p,q\}$ of orthoprojections one then especially
has $v^*v=p$ and $vv^*=q$, that is, $p\sim q$ within $M$. Note that by local invertibility of
the moduli from the uniqueness of the polar decomposition then especially also
$v=qp|qp|^{-1}$ follows.
\begin{lemma}\label{pairs}
For given locally invertible $a\geq {\mathbf 0}$ the following are mutually equivalent:
\begin{enumerate}
\item\label{pairs.1}
$(a,b)\in PAIRS(M)\,;$
\item\label{pairs.2}
$\{p,q\}$ with $q=s(b)$ and $p=s(a)$ is a minimal pair of orthoprojections and
\begin{equation}\label{fi.4}
b=qp|qp|^{-2} a^{-1}|qp|^{-2} pq=q|qp|^{-2} a^{-1}|qp|^{-2} q
\end{equation}
has to be fulfilled.
\end{enumerate}
\end{lemma}
\begin{proof}
Suppose \eqref{pairs.1}. Since $a,b$ are locally invertible from $aba=a$ the relation
$pbp=a^{-1}$ follows, with $p=s(a)$. Let $q=s(b)$. Since $b\leq \lambda q$ holds,
for some $\lambda>0$, we infer that $a^{-1}\leq pbp\leq \lambda pqp\leq \lambda p$.
Hence, $pqp$ has to be locally invertible as well, with $s(pqp)=p$. Analogously,
starting from $bab=b$ under the supposition of \eqref{pairs.1} will show that $qpq$
is locally invertible, with $s(qpq)=q$. Hence $\{p,q\}$ is a minimal pair of orthoprojections.
To see that \eqref{fi.4} is fulfilled, note that in view of the above and with $v=v(p,q)$ we have
$a^{-1}=pbp=pqbqp=|qp|v^*bv|qp|$. Owing to the local invertibility of $|qp|$ and since $vv^*=q$ holds
from this $b=v|qp|^{-1}a^{-1}|qp|^{-1}v^*$ follows. From this in view of $v=qp|qp|^{-1}$ the
formula \eqref{fi.4} is obtained. Thus the implication \eqref{pairs.1}\,$\Longrightarrow$\,\eqref{pairs.2} is
seen to hold.

On the other hand, if in line with item \eqref{pairs.2} and formula
\eqref{fi.4} a minimal pair $\{p,q\}$ of orthoprojections and
locally invertible $a\geq {\mathbf 0}$ with $p=s(a)$ are given, then with $b$ given in accordance with
formula \eqref{fi.4} the following is fulfilled:\ $aba=aqp|qp|^{-2} a^{-1}|qp|^{-2} pqa=
apqp|qp|^{-2} a^{-1}|qp|^{-2} pqpa=apa^{-1}pa=a$. Also,
from $v=qp|qp|^{-1}$ with $vv^*=q$, $v^*v=p$ and \eqref{fi.4} we infer that $s(b)=q$ holds and that
$v^*bv=|qp|^{-1}a^{-1}|qp|^{-1}$ is locally invertible, with local inverse $(v^*bv)^{-1}=|qp| a |qp|$.
Hence, $b$ must be locally invertible, with $b^{-1}=v(v^*bv)^{-1}v^*=qp a pq$. From this one easily infers that
then $a=pq |pq|^{-2} b^{-1}  |pq|^{-2}qp$ holds, that is, also the dual formula to \eqref{fi.4}
has to be fulfilled. Thus analogously also $bab=b$ can be followed (see above), and therefore the implication
\eqref{pairs.2}\,$\Longrightarrow$\,\eqref{pairs.1} holds.
\end{proof}
It is plain to see that from Lemma \ref{pairs} and for
given orthoprojection $r\in M$ the following parametrization of $PAIRS_r(M)$ can be obtained:
\begin{equation}\label{pairs.3}
PAIRS_r(M)=\biggl\{(a,q|qp|^{-2} a^{-1}|qp|^{-2} q):a\geq {\mathbf 0},\{p,q\},\,s(a)=p\approx q\approx r\biggr\}\,,
\end{equation}
where $\{p,q\}$ and $a$ are thought to be minimal pairs of orthoprojections and
locally invertible operators which are subject to the given conditions, accordingly.

\subsection*{Another formula for partial fidelity (general case)}
Let $r\in M\backslash \{{\mathbf 0}\}$ be an orthoprojection.
With the help of \eqref{pairs.3} the defining formula
of the partial fidelity $F_M(\omega,\varrho|r)$ can be turned into the following expression:

\begin{equation*}
F_M(\omega,\varrho|r)=\inf_{\{p,q\},\ p\approx q\approx r}\biggl\{ \inf_{a\in pMp_+,\text{invertible}}\frac{1}{2}\{\omega(a)+\varrho(q|qp|^{-2} a^{-1}|qp|^{-2} q)\}\biggr\}\,,
\end{equation*}
where $\{p,q\}$ is to extend about minimal pairs of orthoprojections. With the help of \eqref{fi.1}
we see that the inner infimum equals
$$F_{pMp}(\omega|_{pMp},\varrho^{|qp|^{-2}q}|_{pMp})=\inf_{a\in pMp_+,\,\text{invertible}}\frac{1}{2}\{\omega(a)+\varrho(q|qp|^{-2} a^{-1}|qp|^{-2} q)\}\,.$$
According to \eqref{fi.3} this is the same as $F_M(\omega^p,\varrho^{|qp|^{-2}q})$.
Owing to $p(|qp|^{-2}q)=(|qp|^{-2})q$ and \eqref{fi.2} we see that
$F_M(\omega^p,\varrho^{|qp|^{-2}q})=F_M(\omega^{|qp|^{-2}},\varrho^q)$.
In summarizing from the previous we thus obtain the following formula:
\begin{subequations}\label{pafi1}
\begin{equation}\label{pafi.2}
F_M(\omega,\varrho|r)=\inf_{\{p,q\},\,p\approx q\approx r} F_M(\omega^{|qp|^{-2}},\varrho^q)\,,
\end{equation}
with the infimum taken over all minimal pairs of orthoprojections which are subject to the
condition $p\approx q\approx r$. Note that it is mainly due to this formula that
there is some hereditary relationship pointing from properties of
fidelity to those of partial fidelity and vice versa. As an example, note that in view
of \eqref{fi.1a} and
\eqref{fi.3}, and since each pair $\{p,q\}$ with $q=p$ and $p\approx r$ is a
special minimal pair of orthoprojections, from \eqref{pafi.2} the following upper
bound can be inferred:
\begin{equation}\label{esti0.1}
F_M(\omega,\varrho|r)\leq \inf_{p\approx r} F_M(\omega^p,\varrho^p)\leq  \inf_{p\approx r}
\sqrt{\omega(p)\varrho(p)}\,.
\end{equation}
Also, with the help of
\eqref{pairs.0} from Definition \ref{pafi} it follows that property \eqref{fi.2}
of fidelity at least for invertible $a,b,c,d$ can be seen to hold with partial fidelity\,:
\begin{equation}\label{pafi.3}
a,b,c,d\in M,\,\text{invertible, with }a^*b=c^*d\,:\
F_M(\omega^a,\varrho^b|r)=F_M(\omega^c,\varrho^d|r)\,.
\end{equation}
\end{subequations}
The formula \eqref{pafi.2} as well as the property \eqref{pafi.3} will be basically for all that follows. Formula \eqref{pafi.2} will be taken as starting point for analyzing the notion of
partial fidelity in more detail. Thereby, note that under the supposition that
${\mathcal H}$ is separable, there is no need for discussing the
properties of partial fidelity in full detail for a general $vN$-algebra. In fact,
under that supposition and at least for normal positive linear forms, with the help of
decomposition theory also the non-factorial cases can be dealt with easily.
This is mainly due to the simple fact that, provided if the system $\{z_k:k\leq n\}$ is a
finite decomposition
of ${\mathbf 1}$ into mutually orthogonal central orthoprojections,
then partial fidelities on each pair $\{\omega,\varrho\}$ of positive
linear forms behave additively:
\begin{equation}\label{esti0.2}
F_M(\omega,\varrho|r)=\sum_{k\leq n} F_{Mz_k}(\omega|_{Mz_k},\varrho|_{Mz_k}|rz_k)\,,
\end{equation}
which fact is a direct consequence from the definition of
$PAIRS_r(M)$ in \eqref{fundef}, Definition \ref{pafi}, \eqref{fi.3} and Lemma \ref{pairs}. It is
obvious that for centrally normal positive linear forms \eqref{esti0.2} remains true for
any (also non-finite) decompositions of ${\mathbf 1}$. Thus, in particular the latter
happens to be true for normal positive linear forms, and as mentioned above then the
general case can be reduced to the factorial cases, by known standard procedures
\cite[{\em Chapter I\,}]{Schw:67}.
\subsection*{Partial fidelities on factors (generalities)}
Throughout this subsection $M$ be a factor on a separable Hilbert space. Thus, there are
the {\em finite} factors of types ${\mathrm I}_n$, $n\in {\mathbb N}$, and type ${\mathrm II}_1$,
the {\em semifinite} factors
of type ${\mathrm I}_\infty$ and type ${\mathrm II}_\infty$, and the {\em purely infinite}
factors of type ${\mathrm III}$, with their respective ranges of the
{\em relative dimension function}. Remind that in the latter case for nonzero
orthoprojections $p,q$ one has always $p\sim q$. In line with this,
as range of the relative dimension function the two-point set $\{0,\infty\}$ will be taken,
where zero and $\infty$ correspond
to the relativ dimensions of ${\mathbf 0}$ and of each $p\not={\mathbf 0}$, respectively.
Therefore, there are then exactly three
classes of mutually unitary equivalent orthoprojections: $\{{\mathbf 0}\}$, $\{p:{\mathbf 0}<p<
{\mathbf 1}\}$ and $\{{\mathbf 1}\}$.
If $M$ is a finite or semifinite factor, then
a nonvanishing normal trace $\tau$ exists and is faithful, and is unique up
to a positive multiple. We fix such a normal faithful trace. Then the relative
dimension of an orthoprojection $p$ may be identified
with $\tau(p)$. For orthoprojections $q,p$ one then
has $p\succ q$ if, and only if, $\tau(p)\geq \tau(q)$. Thus especially $p\sim q$ iff
$\tau(p)=\tau(q)$.

In a factor, refer to the relative dimension of $p^\perp$ as
{\em relative codimension} of $p$. Therefore, $p\approx q$ iff both,
the relative dimensions and the relative codimensions of $p$ and $q$ agree.
In the semifinite case and for projections
$p,q$ with finite relative dimension
(resp.~finite relative codimension) this even simplifies: $q\approx p$ occurs if, and only if,
$\tau(p)=\tau(q)<\infty$
(resp.~$\tau(q^\perp)=\tau(p^\perp)<\infty$).

Start with asking for the scale of orthoprojections $r$ where partial fidelity can
be nontrivial at all. The answer can be obtained with the help of
\eqref{esti0.1}, essentially, and tells us that $F_M(\cdot,\cdot |r)$
yields classes of unitary invariants
which can be labelled through the finite part of the range of the relative (co)dimension function.
\begin{lemma}\label{codim}
Let $M$ be an infinite factor. Then $F_M(\cdot,\cdot |r)\equiv 0$ for each $r$ of infinite relative codimension. In particular, in the purely infinite case, partial fidelity is identically
vanishing unless $r={\mathbf 1}$.
\end{lemma}
\begin{proof}
Suppose $M$ to be infinite, and assume $r$ has infinite relative codimension.
In our context this means that $r^\perp$ is an orthoprojection of
infinite relative dimension. Since we are on a separable Hilbert space, there exists
only one equivalence class of infinite orthoprojections, and thus
$r^\perp\sim {\mathbf 1}$ is properly infinite
(here and henceforth
for operator algebraic notions and details
the reader is referred to \cite{Saka:71} e.g.). Then there is
an infinite system $\{p_n\}$ of mutually orthogonal orthoprojections with $p_n<r^\perp$ and
$p_n\sim r^\perp$, for each $n\in {\mathbb N}$. Note that then also
$p_n\sim r^\perp-p_n\sim r^\perp\sim p_n^\perp$ holds. Let $\omega,\varrho\in M_+^*$.
Then especially $\sum_{n\leq N}\omega(p_n)\leq \omega({\mathbf 1})$ holds, for each
$N\in {\mathbb N}$, and therefore
$\lim_{n\to\infty} \omega(p_n)=0$. Hence,
for given $\varepsilon>0$ and given positive linear forms
$\omega$ and $\varrho$, there exists an orthoprojection $p$ obeying
$p<r^\perp$, $p\sim r^\perp-p\sim r^\perp\sim p^\perp$ and $\omega(p)<\varepsilon$ (take $p=p_n$, for
$n\in {\mathbb N}$ sufficiently large).  Since in the factor case
all orthoprojections are mutually comparable and $p$ is infinite, one has $p\succ r$. Hence,
there is an orthoprojection $s$ obeying
$r\sim s\leq p<r^\perp$. Owing to $s^\perp\geq p^\perp$ and since $p^\perp$ is infinite
also $s^\perp$ is infinite, in which case then also $s^\perp\sim r^\perp$ follows.
Hence, $s\approx r$ and $\omega(s)\leq \omega(p)<\varepsilon$, and thus in accordance with \eqref{esti0.1} one arrives at $F_M(\omega,\varrho|r)\leq
\sqrt{\varepsilon}\sqrt{\varrho({\mathbf 1})}$.
These arguments apply for each $\varepsilon>0$. Hence $F_M(\omega,\varrho|r)=0$ must be fulfilled
whenever $r$ has nonfinite relative codimension.
Since both $\omega,\,\varrho$ can be chosen at will, we have
$F_M(\cdot,\cdot |r)\equiv 0$. Finally, if $M$ is of type ${\mathrm III}$, from the just proven
$F_M(\cdot,\cdot |r)\equiv 0$ follows, for each orthoprojection $r<{\mathbf 1}$. On the other hand,
for $r={\mathbf 1}$ and each $\omega,\varrho\in M_+^*$ one has $F_M(\cdot,\cdot |{\mathbf 1})=
\sqrt{P_M(\omega,\varrho)}$. But one knows that $P_M(\omega,\varrho)=0$ is equivalent
with orthogonality of $\omega$ and $\varrho$, see e.g.~in \cite{AlUh:84}. Thus certainly
$F_M(\cdot,\cdot |{\mathbf 1})\not\equiv 0$.
\end{proof}
In view of Lemma \ref{codim} in the following and for infinite $M$
we must not care about the purely infinite case,
and in the other infinite cases only partial fidelities labelled by equivalence classes of
orthoprojections with finite codimension will be of interest.
\begin{remark}\label{pur}
Note that according to decomposition theory in the purely infinite case
with normal positive linear forms the fact mentioned on in Lemma \ref{codim} straightforwardly
extends from
the purely infinite factorial cases to the nonfactorial cases. That is, for normal
$\omega$ and $\varrho$ one has $F_M(\omega,\varrho|r)=0$ unless $r\not={\mathbf 1}$.
\end{remark}
In order to proceed with
formula \eqref{pafi.2} for infinite $M$, we have to learn something about minimal pairs of
orthoprojections of finite relative codimension first. Start with an auxiliary result, which
makes sense also in the nonfactorial cases.
\begin{lemma}\label{pairsp}
Let $M$ be semifinite, properly infinite, and be $r\in M$
an orthoprojection with finite $r^\perp$.
Let $\{p,q\}$ be a minimal pair, with $p\approx r\approx q$. Then,
there exists a finite orthoprojection $Q$ and
a minimal pair $\{p',q'\}$ of orthoprojections obeying $p',q'\leq Q$, $p'\prec r^\perp$,
$q'\prec r^\perp$ and $p=p'+Q^\perp$ and $q=q'+Q^\perp$.
\end{lemma}
\begin{proof}
By assumption $p,q,r$ all are
infinite and obey $p^\perp\approx q^\perp\approx r^\perp$.
Define $p'=s(pq^\perp p)$. Then $p'\leq p$ and $r^\perp\approx q^\perp \succ p'$.
Hence $r^\perp \succ p'$, and therefore $p'$ has to be finite. Define $Q=(p-p')^\perp$.
Then, from $pqp=p-pq^\perp p$ we get $Q^\perp q Q^\perp=Q^\perp$, which is the same as
$Q^\perp q^\perp Q^\perp={\mathbf 0}$. Hence $q^\perp Q^\perp={\mathbf 0}$, and therefore $Q^\perp\leq q$.
Define $q'=q-Q^\perp$. Then $q'\perp Q^\perp$, and
$p=Q^\perp+p'$ and $q=Q^\perp+q'$ are orthogonal decompositions. It is then also easily inferred that
$pqp=Q^\perp+p'qp'=Q^\perp+p'q'p'$ and $qpq=Q^\perp+q'pq'=Q^\perp+q'p'q'$ must be fulfilled.
Then, from the previous in view of
orthogonality of $Q^\perp$ to $p'q'p'$ and $q'p'q'$ and by minimality of the
pair $\{p,q\}$ we infer that both $p'q'p'$ and $q'p'q'$ must be locally invertible, and have supports
$p'$ and $q'$, respectively. Note from the preceding that also
$Q=q'+q^\perp=p'+p^\perp$ must be fulfilled. Owing to the finiteness of $p',q^\perp,p^\perp$
it follows that also $q'$ and $Q$ have to be finite. Finally, since $QMQ$ is
a finite $vN$-algebra and $q'+q^\perp$ and $p'+p^\perp$ both are orthogonal decompositions of the
corresponding unit $Q$, from $p^\perp\approx q^\perp$ we then can even conclude that
$p'\approx q'$.
\end{proof}
\begin{corolla}\label{facco}
Let $M$ be a semifinite factor, with faithful normal trace $\tau$.
For given minimal
pair $\{p,q\}$ obeying
$p\approx q\approx r$ with finite $r^\perp$, there exists an ascendingly directed sequence
$\{Q_n\}$ of finite orthoprojections with $\bigvee_n Q_n={\mathbf 1}$, and a sequence of
minimal pairs $\{p_n,q_n\}$ obeying $p_n,q_n\leq Q_n$, $\tau(p_n)=\tau(q_n)=\tau(Q_n)-\tau(r^\perp)$,
$p=p_n+Q_n^\perp$, $q=q_n+Q_n^\perp$ and
$\|\,|q_n p_n|^{-2}\|=\|\,|q_1 p_1|^{-2}\|$, for each $n\in {\mathbb N}$.
\end{corolla}
\begin{proof}
Let $Q,p',q'$ be chosen in accordance with the hypothesis of Lemma \ref{pairsp}.
Then $p'=Q-p^\perp$ and $q'=Q-q^\perp$,
with $p^\perp\approx q^\perp\approx r^\perp$. Since $Q$, $p'$ and $q'$ are
finite and $Q^\perp$ is infinite and $M$ is a (countably decomposable)
factor, one has
$\tau(p')=\tau(q')=\tau(Q)-\tau(r^\perp)$, and
there has to exist an ascendingly directed sequence $\{z_n\}$ of finite
orthoprojections $z_n\leq Q^\perp$ with
$\bigvee_n z_n=Q^\perp$. We now define finite orthoprojections by $Q_n=Q+z_n$,
$p_n=p'+z_n$ and $q_n=q'+z_n$, respectively. By assumption on $Q,p',q'$ and construction of
these orthoprojections
it is easily seen that each $\{p_n,q_n\}$ is a minimal pair of orthoprojections,
with $p_n,q_n\leq Q_n$, and which obeys $p=p'+Q^\perp=p_n+Q^\perp-z_n=p_n+Q_n^\perp$, $q=q_n+Q_n^\perp$
and $\tau(p_n)=\tau(q_n)=\tau(Q)-\tau(r^\perp)+\tau(z_n)=\tau(Q_n)-\tau(r^\perp)<\infty$.
Since $M$ is a factor, from this also $p_n\approx q_n$ is seen to hold, for each $n\in {\mathbb N}$.
Finally, note that by construction $|q_np_n|^{-2}=|q'p'|^{-2}+z_n$, with $|q'p'|^{-2}\perp z_n$.
Since $|q'p'|^{-1}$ has norm larger than one, from this $\|\,|q_n p_n|^{-2}\|=\|\,|q' p'|^{-2}\|$
follows, for each $n\in {\mathbb N}$.
\end{proof}
The following result states that semifinite factor cases of partial fidelity at pairs of normal positive linear forms can be reduced to consideration of partial fidelities over finite factors,
essentially.
\begin{satz}\label{ap}
Let $M$ be a semifinite factor and be $r\in M$ an
orthoprojection of finite relative codimension. For each two
normal positive linear forms $\omega,\varrho\in M_{*+}$ and $\varepsilon>0$ there exists
an orthoprojection $p\approx r$ and an ascending sequence $\{Q_n\}$ of finite orthoprojections
with $\bigvee_n Q_n={\mathbf 1}$ and $p\geq Q_k^\perp$, for each $k\in {\mathbb N}$, such that
the following estimates hold\textup{:}
\begin{enumerate}
\item\label{ap1}
$F_M(\omega,\varrho|r)\geq \limsup_{n\to\infty}
F_{Q_nMQ_n}\bigl(\omega|_{Q_nMQ_n},\varrho|_{Q_nMQ_n}\big|\,p-Q_n^\perp\bigr)-\varepsilon\,;$
\item\label{ap2}
$F_M(\omega,\varrho|r)\leq \liminf_{n\to\infty}
F_{Q_nMQ_n}\bigl(\omega|_{Q_nMQ_n},\varrho|_{Q_nMQ_n}\big|\,p-Q_n^\perp\bigr)\,.$
\end{enumerate}
Also, there exist sequences of orthoprojections $\{P_n\}$ and $\{E_n\}$, with $E_n$ finite and
$r\approx P_n\geq E_n^\perp$, for each $n\in {\mathbb N}$, and $\lim_{n\to\infty} E_n={\mathbf 1}$
strongly, such that
\begin{enumerate}
\setcounter{enumi}{2}
\item\label{ap3}
$F_M(\omega,\varrho|r)= \lim_{n\to\infty}
F_{E_nME_n}\bigl(\omega|_{E_nME_n},\varrho|_{E_nME_n}\big|\,P_n-E_n^\perp\bigr)\,.$
\end{enumerate}
\end{satz}
\begin{proof}
In view of \eqref{pafi.2}, a minimal pair $\{p,q\}$ obeying
$p\approx q\approx r$ and
\begin{subequations}\label{cs}
\begin{equation}\label{cs00}
F_M(\omega,\varrho|r)+\frac{\varepsilon}{2} \geq
F_M(\omega^{|qp|^{-2}},\varrho^q)\geq F_M(\omega,\varrho|r)
\end{equation}
can be chosen. Let $\tau$ be a fixed faithful normal trace on $M$.
Since $r^\perp$ is finite, Corollary \ref{facco} can be applied, and provides existence
of an ascendingly directed sequence
$\{Q_n\}$ of finite orthoprojections with $\bigvee_n Q_n={\mathbf 1}$, and a sequence of
minimal pairs $\{p_n,q_n\}$ obeying $p_n,q_n\leq Q_n$, $\tau(p_n)=\tau(q_n)=\tau(Q_n)-\tau(r^\perp)$,
$p=p_n+Q_n^\perp$ and $q=q_n+Q_n^\perp$. Hence $p_n=p-Q_n^\perp$, which means that $p\geq Q_n^\perp$, for each $n\in {\mathbb N}$.
Owing to \eqref{fi.1} invertible
$a>{\mathbf 0}$ exists such that,
for each $n\in {\mathbb N}$, the following holds:
\begin{equation}\label{cs10}
\begin{split}
F_M(\omega^{|qp|^{-2}},\varrho^q)+\frac{\varepsilon}{2}& > \frac{1}{2} \bigl\{\omega^{|qp|^{-2}}(a)+
\varrho^{q}(a^{-1})\bigr\}\\
& \geq \frac{1}{2}\bigl\{\omega^{|q_np_n|^{-2}}(a)+
\varrho^{q_n}(a^{-1})\bigr\}+ R_n\\
& \geq F_M(\omega^{|q_np_n|^{-2}},\varrho^{q_n})+ R_n\,,
\end{split}
\end{equation}
with the remainder $R_n=\Re\bigl\{\omega(|q_np_n|^{-2}aQ_n^\perp)+
\varrho(q_n a^{-1}Q_n^\perp)\bigr\}$. Note that according to Corollary \ref{facco} the
sequence $\{|q_np_n|^{-2}\}$ is uniformly bounded by $\bigl\| |q_1p_1|^{-2}\bigr\|$,
and $\{Q_n^\perp\}$ is strongly tending to
zero as $n$ tends to infinity. Thus, by normality and since
$\bigl|\omega(|q_np_n|^{-2}aQ_n^\perp)\bigr|^2\leq\omega(|q_np_n|^{-2}a^2|q_np_n|^{-2})
\omega(Q_n^\perp)\leq \|a\|^2\bigl\| |q_1p_1|^{-2}\bigr\|^2\omega({\mathbf 1})\omega(Q_n^\perp)$ and
$|\varrho(q_n a^{-1}Q_n^\perp)|^2\leq \varrho(q_na^{-2}q_n)\varrho(Q_n^\perp)\leq \|a^{-2}\|
\varrho({\mathbf 1})\varrho(Q_n^\perp)$ hold, the relation $\lim_{n\to\infty} R_n=0$ can be
inferred. On the other hand, owing to \eqref{fi.3} and \eqref{pafi.2}, and since $q_n,p_n\leq Q_n$ and $\tau(p_n)=\tau(q_n)=\tau(Q_n)-\tau(r^\perp)=\tau(Q_n)-\tau(p^\perp)=\tau(Q_n-p^\perp)=\tau(p-Q_n^\perp)<\infty$ hold, the following estimate is seen to hold:
\begin{equation*}
\begin{split}
F_M(\omega^{|q_np_n|^{-2}},\varrho^{q_n})& =
F_{Q_nMQ_n}\bigl(\bigl(\omega|_{Q_nMQ_n}\bigl)^{|q_np_n|^{-2}},
\bigl(\varrho|_{Q_nMQ_n}\bigr)^{q_n}\bigr)\\
& \geq
F_{Q_nMQ_n}\bigl(\omega|_{Q_nMQ_n},
\varrho|_{Q_nMQ_n}\big|\,p-Q_n^\perp\bigr)\,.
\end{split}
\end{equation*}
In summarizing, from \eqref{cs10} together with the latter estimates and the fact that $\lim_{n\to\infty} R_n=0$ one concludes that
$$
F_M(\omega^{|qp|^{-2}},\varrho^q)+\frac{\varepsilon}{2}\geq \limsup_{n\to\infty}F_{Q_nMQ_n}\bigl(\omega|_{Q_nMQ_n},
\varrho|_{Q_nMQ_n}\big|\,p-Q_n^\perp\bigr)
$$
must be fulfilled. From the latter and \eqref{cs00} the validity of \eqref{ap1} follows.

To see that \eqref{ap2} is true, remark first that if $\{p',q'\}$ is a minimal pair with
$p',q'\leq Q_n$
and $\tau(p')=\tau(p_n)=\tau(Q_n)-\tau(r^\perp)=\tau(q')$, then $\{p'+Q_n^\perp,q'+Q_n^\perp\}$ will
be another minimal pair
of orthoprojections with $\tau((p'+Q_n^\perp)^\perp)=\tau(({\mathbf 1}-p'-Q_n^\perp))=
\tau((Q_n-p'))=\tau(Q_n)-\tau(p')=\tau(r^\perp)<\infty$, and analogously $\tau((q'+Q_n^\perp)^\perp)=\tau(r^\perp)<\infty$. Hence, $p'+Q_n^\perp\approx q'+Q_n^\perp\approx r$.
Thus, in view of \eqref{pafi.2} we have
\begin{equation*}
F_M(\omega,\varrho|r)
\leq \inf_{\{p',q'\},\,p',q'\leq Q_n,\,\tau(p')=\tau(Q_n)-\tau(r^\perp)}
F_M(\omega^{|q'p'|^{-2}+Q_n^\perp},\varrho^{q'+Q_n^\perp})\,.
\end{equation*}
Applying \eqref{fi.2'} under the infimum and using \eqref{fi.3} and \eqref{pafi.3} once more
again then yields the following (remind that $p_n=p-Q_n^\perp$):
$$
F_M(\omega,\varrho|r)\leq
 \inf_{\{p',q'\},\,p',q'\leq Q_n,\,\tau(p')=\tau(Q_n)-\tau(r^\perp)}
F_M(\omega^{|q'p'|^{-2}},\varrho^{q'})+F_M(\omega^{Q_n^\perp},\varrho^{Q_n^\perp})\,,
$$
and therefore one has
\begin{equation}\label{cs13}
F_M(\omega,\varrho|r)\leq F_{Q_nMQ_n}\bigl(\omega|_{Q_nMQ_n},
\varrho|_{Q_nMQ_n}\big|\,p-Q_n^\perp\bigr)+F_M(\omega^{Q_n^\perp},\varrho^{Q_n^\perp})\,.
\end{equation}
Note that from \eqref{esti0.1}, $F_M(\omega^{Q_n^\perp},\varrho^{Q_n^\perp})\leq
\sqrt{\omega(Q_n^\perp)\varrho(Q_n^\perp)}$ can be seen. Thus, by normality of the
positive linear forms, $\lim_{n\to\infty}F_M(\omega^{Q_n^\perp},\varrho^{Q_n^\perp})=0$
follows, and thus then
$F_M(\omega,\varrho|r)\leq\liminf_{n\to\infty}F_{Q_nMQ_n}\bigl(\omega|_{Q_nMQ_n},
\varrho|_{Q_nMQ_n}\big|\,p-Q_n^\perp\bigr)$
is seen to hold, and thus \eqref{ap2} is valid.

To see \eqref{ap3}, let $p=p^{(k)}\approx r$ and the ascending sequence of finite orthoprojections
$\{Q_n\}=\{Q_n^{(k)}\}$ with $\bigvee_n Q_n^{(k)}={\mathbf 1}$ be chosen as to satisfy
\eqref{ap1} and \eqref{ap2},
to given $\omega$ and $\varrho$, but with $\varepsilon=1/k$.
Let $F=F_M(\omega,\varrho|r)$, and be $F_n^{(k)}$ defined accordingly, but referring to
$Q_n^{(k)}MQ_n^{(k)}$ and the restrictions of $\omega$ and $\varrho$ to this subalgebra, and
$p^{(k)}-{Q_n^{(k)}}^\perp$ instead of $r$. Then, these assumptions in view of \eqref{ap1} and
 \eqref{ap2} read as
\begin{equation}\label{cs11}
\liminf_{n\to\infty} F_n^{(k)}\geq F\geq \limsup_{n\to\infty} F_n^{(k)} +\frac{1}{k}\,,
\end{equation}
for each $k\in {\mathbb N}$. Note that from this especially also follows that both
$\limsup_{n\to\infty} F_n^{(k)}$ and $\liminf_{n\to\infty} F_n^{(k)}$ are finite, for each
$k\in {\mathbb N}$.
Let $\{\varphi_n\}$ be a complete
orthonormal system in ${\mathcal H}$ (remind that $M$ is acting on a {\em separable} Hilbert space).
We then chose an ascending sequence of subscripts $\{n_k\}$ inductively, by the following procedure:
if $n_{k-1}$ has been yet chosen, choose $n_k>n_{k-1}$ such that $\|Q_{n_k}^{(k)}\varphi_j-\varphi_j\|\leq 1/k$, for all $j\leq k$, and
\begin{equation}\label{cs12}
\limsup_{n\to\infty} F_n^{(k)} +\frac{1}{k}\geq F_{n_k}^{(k)}\geq \liminf_{n\to\infty} F_n^{(k)}
-\frac{1}{k}
\end{equation}
are fulfilled. From \eqref{cs11} and \eqref{cs12} then $F=\lim_{k\to\infty}  F_{n_k}^{(k)}$ follows.
In defining $E_k=Q_{n_k}^{(k)}$ and $P_k=p^{(k)}$, for $k\in {\mathbb N}$, we thus arrive at
\eqref{ap3}. Also, note that by choice of $n_k$, $\lim_{k\to\infty} E_k\varphi_j=\varphi_j$ is
fulfilled, for each $j\in {\mathbb N}$. Hence, $\lim_{k\to\infty} E_k\varphi=\varphi$ for all $\varphi$ of a dense subspace of ${\mathcal H}$. By uniform boundedness of $\{E_k\}$ from this
$\lim_{k\to\infty} E_k\varphi=\varphi$ follows, for each $\varphi\in {\mathcal H}$, that is,
in accordance with the assertion, $\{E_n\}$ is strongly tending to ${\mathbf 1}$.
\end{subequations}
\end{proof}
The conclusions within the proof of Proposition
\ref{ap}\,\eqref{ap2}, and there especially the arguments leading
to \eqref{cs13}, may be applied also in a slightly more general
context. To explain this, note that since $M$ is a {\em factor},
to each orthoprojection $Q$ obeying $\tau(r^\perp)\leq
\tau(Q)<\infty$ another orthoprojection $P$ with $P\leq Q$ exists
and which obeys $\tau(P)=\tau(Q)-\tau(r^\perp)$. The mentioned
arguments then imply the following to hold\textup{:}
\begin{subequations}\label{cscomm}
\begin{equation}\label{cs14}
F_M(\omega,\varrho|r)-\sqrt{\omega(Q^\perp)\varrho(Q^\perp)}\leq
F_{QMQ}\bigl(\omega|_{QMQ},
\varrho|_{QMQ}\big|\,P\bigr)\,.
\end{equation}
In line with this, for any ascending sequence $\{Q_n\}$ of finite orthopojections with
$\tau(Q_n)\geq \tau(r^\perp)$ and $\bigvee_n Q_n={\mathbf 1}$, there exists another
ascending sequence $\{P_n\}$ of orthoprojections, with $P_n\leq Q_n$ and
$\tau(P_n)=\tau(Q_n)-\tau(r^\perp)$,
such that
\begin{equation}\label{cs15}
F_M(\omega,\varrho|r)\leq \liminf_{n\to\infty} F_{Q_nMQ_n}\bigl(\omega|_{Q_nMQ_n},
\varrho|_{Q_nMQ_n}\big|\,P_n\bigr)
\end{equation}
\end{subequations}
must be fulfilled.
\begin{remark}\label{semifi}
According to Proposition \ref{ap}, partial fidelity on a semifinite factor and
for pairs of normal positive linear forms is uniquely determined by the partial fidelities
coming along by restricting arguments to hereditary {\em finite} subfactors, accordingly.
Note that
for fidelity ($r={\mathbf 1}$) this is a well-known fact, which in view of $F_M=\sqrt{P_M}$
at once can be followed from \eqref{fi.2a} and \eqref{fi.3}.
\end{remark}

\subsection*{Traces and partial fidelity}
The most important special cases of $F_M(\omega,\varrho|r)$ to be considered
throughout the rest of the paper
will be if $\tau$ is a trace on $M$ and $\omega$ and
$\varrho$ are inner derived positive linear forms of $\tau$, that is, one has
$\omega=\tau^a$ and $\varrho=\tau^b$, with $a,b\in{\mathcal L}^2(M,\tau)$. In this situation
\eqref{pafi.2} reads as
\begin{subequations}\label{tr}
\begin{equation}\label{tr.1}
F_M(\tau^a,\tau^b|r)=\inf_{\{p,q\},\,p\approx r} F_M(\tau^{|qp|^{-2}a},\tau^{qb})\,.
\end{equation}
In order to evaluate the expession
under the infimum the formula \eqref{fi.7} can be applied, provided $\tau$ is lower semi-continuous, or
if some additional conditions in respect of the supports of $|a|$ and $|b|$ are fulfilled
(cf.~Corollary \ref{centcor}). In all these cases one has
$F_M(\tau^{|qp|^{-2}a},\tau^{qb})=\tau(|a|qp|^{-2}qb|)$, and \eqref{tr.1} turns into
\begin{equation}\label{tr.2}
F_M(\tau^a,\tau^b|r)=\inf_{\{p,q\},\,p\approx r}  \tau(|a^*|qp|^{-2}qb|)\,.
\end{equation}
In view of Corollary \ref{centcor} one knows
that formula \eqref{tr.2} holds for lower semi-continuous $\tau$, and remains true for any trace if
$r\in {\mathcal L}^1(M,\tau)$ holds or if both $s(|a|)$ and $s(|b|)$ belong to
${\mathcal L}^1(M,\tau)$, for instance. For simplicity, in the sequel often we will
suppose $\tau$ to be lower semi-continuous. Also remark that
since $\tau^c=\tau^{|c^*|}$ is fulfilled, for any $c\in {\mathcal L}^2(M,\tau)$, in analyzing partial
fidelity we may content with considering the structure of \eqref{tr.2} in the special case of
positive elements $a,b\in{\mathcal L}^2(M,\tau)_+$. Since partial fidelity
obviously is a symmetric setting with respect to the functional arguments,
one especially must have that \eqref{tr.2} is equivalent to
\begin{equation}\label{tr.2b}
F_M(\tau^a,\tau^b|r)=\inf_{\{p,q\},\,p\approx r}  \tau(|b^*|qp|^{-2}q a|)\,.
\end{equation}
\end{subequations}
Suppose now that $a,b\in{\mathcal L}^2(M,\tau)_+$. Let us define invertible positive operators
$a_n=\sqrt{a^2+(1/n){\mathbf 1}}$ and
$b_n=\sqrt{b^2+(1/n){\mathbf 1}}$, for each $n\in {\mathbb N}$. With the help of property \eqref{pafi.3}
we then easily infer that $F_M(\tau^a,\tau^b|r)=
F_M(\tau^{b_na},\tau^{b_n^{-1}b})=F_M(\tau^{a_n^{-1}a},\tau^{a_nb}|r)$ has to be fulfilled. Applying \eqref{tr.2}
and invariance in case of a lower semi-continuous trace $\tau$ then provides the following auxiliary identities,
which hold for all $n\in {\mathbb N}$\,:
\begin{equation}\label{auxfi}
F_M(\tau^a,\tau^b|r)=\inf_{\{p,q\},\,p\approx r}  \tau(|b b_n^{-1}q|qp|^{-2}b_n a |)=
\inf_{\{p,q\},\,p\approx r}  \tau(|a a_n^{-1}|qp|^{-2}q a_n b |)\,.
\end{equation}
Start with deriving an upper bound
of $F_M(\tau^a,\tau^b|r)$ in some important special case.
\begin{lemma}\label{upper}
Let $\tau$ be a lower semi-continuous trace and $a,b\in{\mathcal L}^2(M,\tau)_+$.
Suppose that $l(qs(b))\in {\mathcal L}^1(M,\tau)$ holds for each orthoprojection $q\approx r$. Then
\begin{subequations}\label{tresti}
\begin{equation}\label{tr.2a}
F_M(\tau^a,\tau^b|r)\leq \inf_{q\approx r} \tau(\sqrt{q \,|a b|^2 q})
\end{equation}
is fulfilled. In case if $r\in {\mathcal L}^1(M,\tau)$ or both $s(a)$ and $s(b)$ belong to
${\mathcal L}^1(M,\tau)$ the previous estimate remains true for any trace. Especially, in case
of a trace $\tau$ obeying ${\mathbf 1}\in {\mathcal L}^1(M,\tau)$ in addition one finds
\begin{equation}\label{tr.2aa}
F_M(\tau^a,\tau^b|r)\geq \inf_{q\approx r} \tau(|a b| q)\,.
\end{equation}
\end{subequations}
\end{lemma}
\begin{proof}
Assume $l(qs(b))\in {\mathcal L}^1(M,\tau)$. Since $l(qs(b))$ is an orthoprojection one also has
$l(qs(b))\in {\mathcal L}^2(M,\tau)$. Let $s=s(b)$, and consider $\{b_n\}$ as introduced in context of \eqref{auxfi}.
For each $n\in {\mathbb N}$ one then has
${\mathbf 0}\leq b b_n^{-1}\leq s$. Hence also $(b b_n^{-1})^2\leq s$. Note that from this and one of
the identities in \eqref{auxfi}
by monotony and operator monotony of the square root and by invariance $F_M(\tau^a,\tau^b|r)\leq
\inf_{\{p,q\},\,p\approx r}  \tau(| s q|qp|^{-2}b_n a |)=\inf_{\{p,q\},\,p\approx r} \tau(|a b_n|qp|^{-2}q s|)$
can be followed. Also, $b_n\to b$ implies $|a b_n|qp|^{-2}q s|\to |a b|qp|^{-2}q s|$, in a uniform sense.
Hence, since
$$\pm \left\{|a b_n|qp|^{-2}q s|-|a b|qp|^{-2}q s|\right\}\leq \left\|\,|a b_n|qp|^{-2}q s|-|a b|qp|^{-2}q s|\,\right\| l(sq)$$
is fulfilled, owing to $l(sq)=r(qs)\sim l(qs)\in{\mathcal L}^1(M,\tau)$ one has $\tau(l(qs))<\infty$, and thus
from the previous $\lim_{n\to\infty} \tau(|a b_n|qp|^{-2}q s|)= \tau(|a b|qp|^{-2}q s|)$ can be inferred. But then
$\limsup_{n\to\infty} \inf_{\{p,q\},\,p\approx r} \tau(|a b_n|qp|^{-2}q s|)\leq
\inf_{\{p,q\},\,p\approx r} \tau(|a b|qp|^{-2}q s|)$. In view to the above estimate we may summarize\,:
\begin{equation}
F_M(\tau^a,\tau^b|r)\leq \inf_{\{p,q\},\,p\approx r} \tau(|a b|qp|^{-2}q s|)\,.
\tag{$\dag$}
\end{equation}
Now, by invariance of $\tau$ and using once more again monotony and operator monotony of the square root, from
$s\leq {\mathbf 1}$ we conclude $ \tau(|a b|qp|^{-2}q s|)= \tau(|s q |qp|^{-2}ba|)\leq \tau(| q |qp|^{-2}ba|)=
\tau(|ab |qp|^{-2}q|)$.
But then we have $\inf_{\{p,q\},\,p\approx r} \tau(|a b|qp|^{-2}q s|)\leq \inf_{\{p,q\},\,p\approx r} \tau(|ab |qp|^{-2}q|)$.
In view of the latter and ($\dag$), and since $p=q\approx r$ is possible as a special choice for a minimal pair of
orthoprojections and therefore $\inf_{\{p,q\},\,p\approx r} \tau(|ab |qp|^{-2}q|)\leq \inf_{q\approx r} \tau(|ab q|)$
must be fulfilled, we then arrive at \eqref{tr.2a}. The validity of the remaining assertions concerning
the validity of \eqref{tr.2a}
follow along the same procedure since according
to Corollary \ref{centcor} under the mentioned additional premises formula \eqref{fi.7} can be even applied
with any trace.

\noindent
To see \eqref{tr.2aa} note that under the premise of $\tau({\mathbf 1})<\infty$ the
trace $\tau$ even is
a tracial positive linear form. Thus \eqref{auxfi} then also remains valid if $b$ formally
is replaced with $b_n$, that is, one has $F_M(\tau^a,\tau^{b_n}|r)=
\inf_{\{p,q\},\,p\approx r}  \tau(|q|qp|^{-2}b_n a |)$, for each $n\in {\mathbb N}$. Since $\{p,q\}$ is
a minimal pair of orthoprojections the simple identity
$|qp|^{-2}q|qp|^{-2}=|qp|^{-2}pqp|qp|^{-2}=|qp|^{-2}|qp|^2 |qp|^{-2}= |qp|^{-2}$ holds, and therefore one also
infers that $|qp|^{-2}q|qp|^{-2}\geq p$ must be fulfilled.
Hence, our usual combination of monotony and operator monotony arguments will apply and yields that
$|q|qp|^{-2}b_n a |\geq |p b_na|$ has to hold. By invariance of $\tau$ from this
$\tau(|q|qp|^{-2}b_n a |)\geq \tau(|a b_n p|)$ is obtained. Now, once more again by
monotony and operator monotony we als have $|a b_n p|\geq \sqrt{p|ab_n|p|ab_n|p}=p|ab_n|p$, and
thus from this by
invariance of $\tau$ we see that $\tau(|a b_n p|)\geq \tau(|ab_n|p)$ must be fulfilled. The conclusion from all
that is $F_M(\tau^a,\tau^{b_n}|r)\geq \inf_{p\approx r} \tau(|ab_n|p)$, for each $n\in {\mathbb N}$. Since $\tau$
now is a positive linear form, we can be assured that $\lim_{n\to\infty} \tau(|ab_n|p)=\tau(|ab|p)$, for each
$p$. Now, let $q_n\approx r$ with $\tau(|ab_n|q_n)\leq \inf_{p\approx r} \tau(|ab_n|p)+(1/n)$.
We then get
\begin{equation*}
\begin{split}
\inf_{p\approx r} \tau(|ab & |p)  -\inf_{p\approx r} \tau(|ab_n|p)  \leq \tau(\{|ab|-|ab_n|\}q_n)+ 1/n\leq \\
 & \leq \sqrt{\tau(r)}\sqrt{\tau(\{|ab|-|ab_n|\}^2)}+1/n \leq
\sqrt{\tau({\mathbf 1})}\sqrt{\tau(\{|ab_n|-|ab|\}^2)}+ 1/n\,,
\end{split}
\end{equation*}
for each $n\in {\mathbb N}$. Since owing to
$\lim_{n\to\infty} |ab_n|=|ab|$ in the case at hand also
$\lim_{n\to\infty}\tau(\{|ab_n|-|ab|\}^2)=0$ follows, from this in view of the previous we conclude
\begin{equation*}
\limsup_{n\to\infty} F_M(\tau^a,\tau^{b_n}|r)\geq \limsup_{n\to\infty}  \tau(|ab_n|p)\geq \inf_{p\approx r} \tau(|ab|p)\,.
\tag{$\dag\dag$}
\end{equation*}
On the other hand, from $b_n^2\geq b^2\geq {\mathbf 0}$ we get $\tau^{b_n}\geq \tau^b$,
with $\lim_{n\to\infty} \tau^{b_n}=\tau^b$. Hence
$F_M(\tau^a,\tau^{b_n}|r)\geq F_M(\tau^a,\tau^b|r)$ holds, and thus from this together with upper
semi-continuity
of $F_M$ we infer that even $\limsup_{n\to\infty} F_M(\tau^a,\tau^{b_n}|r)=F_M(\tau^a,\tau^b|r)$
has to be fulfilled.
In view of ($\dag\dag$) then \eqref{tr.2aa} is seen.
\end{proof}

\subsection*{Partial fidelity on a finite $vN$-algebra}
Within this subsection partial fidelities will be considered more in detail on a
finite $vN$-algebra $M$ and for inner derived positive linear forms $\omega=\tau^a$ and $\varrho=\tau^b$,
where $\tau$ is a trace
on $M$, and $a$ and $b$ obey $a,b\in {\mathcal L}^2(M,\tau)$.

We start with some general results on finite traces on a $vN$-algebra $M$. Remind that a trace $\tau$
is said to be a {\em finite} trace if ${\mathbf 1}\in {\mathcal L}^1(M,\tau)$ is fulfilled, that is, if $\tau$
is a tracial positive linear form. Note that in this case then
$M={\mathcal L}^1(M,\tau)$ is fulfilled, and thus
especially each orthoprojection
$r$ obeys $r\in {\mathcal L}^1(M,\tau)$. Hence, the suppositions for an application of Lemma \ref{upper}
are fulfilled, and therefore both estimates of \eqref{tresti} hold. In line with this, in case of a
{\em finite trace} $\tau$ on a $vN$-algebra $M$ for each orthoprojection
$r\in M$ we have\,:
\begin{equation}\label{aux.4}
\forall\,a,b\in M_+\,:\ \inf_{q\approx r} \tau(|ab|q)\leq
F_M(\tau^a,\tau^b|r)\leq \inf_{p\approx r}
\tau(\sqrt{p|ab|^2p})\,.
\end{equation}
From this the following result is obtained\,:
\begin{satz}\label{pre}
Let $\tau$ be a finite trace on a $vN$-algebra $M$, and $a,b\in M_+$. Suppose that for given
orthoprojection $r\in M$ the infimum $\inf_{q\approx r}  \tau(|ab|q)$ is attained for some $q=q_0\approx r$ with
$[q_0,|ab|]={\mathbf 0}$. Then, the infimum is the value of the $r$-partial fidelity between $\omega=\tau^a$ and
$\varrho=\tau^b$, $F_M(\tau^a,\tau^b|r)=\inf_{q\approx r}  \tau(|ab|q)$.
\end{satz}
\begin{proof}
Assume $q_0\approx r$ with $\inf_{q\approx r}  \tau(|ab|q)=\tau(|ab|q_0)$ and $[q_0,|ab|]=
{\mathbf 0}$. From the latter
$\sqrt{q_0|ab|^2q_0}=\sqrt{q_0|ab|^2}=q_0|ab|$ follows. Hence, $\tau(\sqrt{q_0|ab|^2q_0})=\tau(|ab|q_0)$.
By assumption $\inf_{q\approx r}  \tau(|ab|q)=\tau(|ab|q_0)=\tau(\sqrt{q_0|ab|^2q_0})\geq
\inf_{p\approx r}  \tau(\sqrt{p|ab|^2p})$ is seen then. From this in view of
\eqref{aux.4} the assertion follows.
\end{proof}
\begin{remark}\label{finfac1}
Note that in the case if $M$ is a finite factor then a trace $\tau$ either can be trivial, that is,
$\tau={\mathbf 0}$ or  ${\mathcal L}^1(M,\tau)=\{{\mathbf 0}\}$ is fulfilled, or it is a nonvanishing
{\em finite} trace.
But then, in case of a finite factor $M$ the assertion of Proposition \ref{pre} remains true for each trace
on $M$.
\end{remark}
We will now specify $M$ to be a finite factor and $\tau$ to be normal.
Then one has the following auxiliary result
which sounds quite classically.
\begin{lemma}\label{pre1}
Let $M$ be a finite factor and $z\in M_+$, and be $s$ an orthoprojection. Let $\tau$ be a trace on $M$.
There exists an orthoprojection $s_0\approx s$ which obeys
both $s_0 z=z s_0$ and $\sup_{q\approx s} \tau(zq)=\tau(zs_0)$.
\end{lemma}
\begin{proof}
Note that according to Remark \ref{finfac1}, $\tau$ must even be a finite trace.
To be non-trivial we may suppose that $\tau\not=0$ and $z\not={\mathbf 0}$ hold.
Without loss of generality in the proof for simplicity $\tau$ and $z$ may be also
supposed to obey $\tau({\mathbf 1})=1$ and $\|z\|=1$, respectively.
Also, since $M$ is a finite factor,
there exists a unique {\em normal} trace $\tau_0$ obeying $\tau_0({\mathbf 1})=1$.

By the spectral theorem there exists a
unique spectral representation of $z$ as an operator Stieltjes-integral
$z=\int_0^1 \lambda\,E(d\/\lambda)$,
with projection-valued measure $E(d\/\lambda)$ derived from a left-continuous spectral family
$\{E(\lambda):\lambda\in {\mathbb R}\}$, that is, a family of orthoprojections obeying
$E(t)\leq E(\lambda)$, for $t\leq \lambda$, $E(t)={\mathbf 0}$, for $t\leq 0$ and
$E(\lambda-)=\bigvee_{t<\lambda} E(t)=E(\lambda)$, for each $\lambda\leq 1$, and with
$E(t)={\mathbf 1}$ for $t>1$.
By convention, for $a<b$, then
$\int_a^{b-} E(d\/\lambda)=E(b)-E(a)=E([a,b[\,)$ and
$\int_a^b E(d\/\lambda)=E(b+)-E(a)=E([a,b])$, and so on accordingly,
where e.g.~$E(b+)$ stands for the greatest lower bound
$E(b+)=g.l.b._{t>b} E(t)$.
Let us define a real $\alpha_s(z)\in [0,1]$ by
$\alpha_s(z)=\max \{ \lambda\in {\mathbb R}: \tau_0(E(\lambda)^\perp)\geq \tau_0(s)\}$.
Note that by our suppositions on $\tau_0$ and $\{E(\lambda)\}$ such number $\alpha_s(z)$ must exist.
Also note that then $\tau_0(E(\alpha_s(z)+)^\perp)\leq  \tau_0(s)$ has to be fulfilled.

Define an
orthoprojection $s_0$ as follows\,: we put $s_0=E(\alpha_s(z)+)^\perp+s_1$,
with an arbitrary orthoprojection $s_1$ obeying $s_1\leq \{E(\alpha_s(z))^\perp-E(\alpha_s(z)+)^\perp\}$ and
$\tau_0(s_1)=\tau_0(s)-\tau_0(E(\alpha_s(z)+)^\perp)$. By the previous and
since we are in a finite factor and $\tau_0$ is the standard tracial state
such choice must be possible. But then we can be sure that $\tau_0(s_0)=\tau_0(s)$
is fulfilled. Since we are in a finite factor this also means that $s_0\approx s$.
By construction of $s_0$ we have $E(\lambda)^\perp\leq s_0$, for $\lambda>\alpha_s(z)$, and
$s_0\leq E(\lambda)^\perp$ for $\lambda\leq\alpha_s(z)$. But then especially
$E(\lambda) s_0=s_0 E(\lambda)$, for each $\lambda$, and therefore $s_0$ commutes with $z$.
Note that from this for (each trace) $\tau$ the following is seen:
\begin{equation*}
\forall\,\lambda\in {\mathbb R}\,:\ \tau(E(\lambda)^\perp s_0)=\min\{\tau(s),\tau(E(\lambda)^\perp)\}\,.
\tag{$\times$}
\end{equation*}
Now, since $\tau$ is tracial, for each two orthoprojections $q,p$ we have $\tau(qp)=\tau(pqp)\leq \tau(p)$
and $\tau(qp)=\tau(qpq)\leq \tau(q)$. Hence $\tau(qp)\leq \min\{\tau(q),\tau(p)\}$. Thus in view of
($\times$) we arrive at the following property\,:
\begin{equation*}
\forall\,\lambda\in {\mathbb R}\,:\ \sup_{q\approx s} \tau(E(\lambda)^\perp q)=\min\{\tau(s),\tau(E(\lambda)^\perp)\}=
\tau(E(\lambda)^\perp s_0)\,.
\tag{$\times\times$}
\end{equation*}
Also, since $\tau$ is a positive linear form (so is bounded), with the help of monotone uniform approximations
of $z$ from below by means of operators $z_n\leq z$, which all belong to the commutative
$vN$-algebra generated by $z$ and which have spectrum consisting of finitely many values,
it can be easily derived that $\tau(z q)$ finally can be represented in Stieltjes form as
$$\tau(z q)=\int_0^{1+} d\/\lambda\, \tau(E(\lambda)^\perp q)$$ (more precisely,
$d\/\lambda$ should be referred to as $f(d\/\lambda)$, with the identity map $f(\lambda)=\lambda$ over the
real axis). From this in view of ($\times\times$) then
$$\sup_{q\approx s} \tau(z q)\leq \int_0^{1+} d\/\lambda\, \sup_{q\approx s} \tau(E(\lambda)^\perp q)=
\int_0^{1+} d\/\lambda\, \tau(E(\lambda)^\perp s_0)=\tau(z s_0)$$ can be followed. This is the same
as  $\sup_{q\approx s} \tau(zq)=\tau(zs_0)$.
\end{proof}
Note that in the finite dimensional case the previous result is well-known and at once follows from
\cite{KyFa:51} and since the $s$-orbits $q\approx s$ are compact.
\begin{remark}\label{trnorm}
Suppose $M$ is a finite $vN$-algebra and $\tau$ is a trace on $M$ which is {\em normal} in restriction
to the center $Z$ of $M$, that is, $\tau|_{Z}$ be normal.
Then, the question on existence of an orthoprojection $s_0\approx s$ which obeys
both $s_0 z=z s_0$ and $\sup_{q\approx s} \tau(zq)=\tau(zs_0)$ in a straightforward manner via
central decomposition of $M$ can be reduced to the same question over finite factors and
traces there (we omit any details on that). Since under the latter condition we have
Lemma \ref{pre1}, in view of the structure of the assertion of Lemma \ref{pre1} and the nature of the
central decomposition and by the central normality of $\tau$ we therefore may even conclude that the conclusion of
Lemma \ref{pre1} can be extended to hold true under these circumstances also in the non-factorial
case of a finite $vN$-algebra. Note that this argument especially yields that
the hypothesis of Lemma \ref{pre1} remains true for each normal trace on a finite $vN$-algebra.
\end{remark}
Now we are ready to end up with our main result for finite $vN$-algebras.
\begin{satz}\label{finfid}
Let $M$ be finite, and be $\tau$ a normal trace. Let $x,y\in {\mathcal L}^2(M,\tau)$, and
$r\in M$, orthoprojection. Then $
F_M(\tau^x,\tau^y|r)=
\inf_{q\approx r}  \tau(|x^*y|q)=\min_{q\approx r}  \tau(|x^*y|q)$.
\end{satz}
\begin{proof}
Before going into the details of the proof of the formula, first we are going
to simplify the problem. By traciality of $\tau$ we have
$\tau^x=\tau^a$ and $\tau^y=\tau^b$, with non-negative operators
$a=|x^*|$ and $b=|y^*|$ which both belong to ${\mathcal
L}^2(M,\tau)_+$. Let $x=w|x|$ be the polar decomposition of $x$.
Then $w^*w=s(|x|)$ and $ww^*=s(|x^*|)$, and $w|x|w^*=|x^*|$ is
equivalent to $x=|x^*|w$. In a finite $vN$-algebra, for
orthoprojections $q,p$ one has that $q\sim p$ implies $q^\perp\sim
p^\perp$, and thus each partial isometry can be extended to an
unitary element in $M$. Thus equivalence in the $vN$-sense is the
same as unitary equivalence in $M$. In line with these facts there
is unitary $u\in M$ such that $w=us(|x|)=s(|x^*|)u$,
i.e.~$x=|x^*|u$, and then $|x^*y|=u^*\sqrt{|y^*|xx^*|y^*|}u=
u^*|\,|x^*|\cdot|y^*|\,|u=u^*|ab|u$ implies that $\inf_{q\approx r}
\tau(|x^*y|q)=\inf_{q\approx r} \tau(|ab|q)$, and if the one infimum
is attained then certainly also the other one will be attained.
According to this, the formula in question becomes equivalent to
\begin{equation*}
F_M(\tau^a,\tau^b|r)=
\inf_{q\approx r}  \tau(|ab|q)=\min_{q\approx r}  \tau(|ab|q)\,,
\tag{$*$}
\end{equation*}
with $a,b\in {\mathcal L}^2(M,\tau)_+$. First, we are going to
prove ($*$) under the additional supposition that $M$ be a {\em
finite factor}. Apply Lemma \ref{pre1} with $z=|ab|$ and
$s=r^\perp$. According to this there exists $s_0\approx s$ which
commutes with $|z|$ and obeys $\sup_{p\approx s}
\tau(zp)=\tau(zs_0)$. Once more again using that in a finite
$vN$-algebra $q\sim s$ is equivalent to $q^\perp\sim s^\perp$,
and vice versa, we have $\inf_{q\approx r}  \tau(z
q)=\tau(z)-\sup_{q^\perp\approx s}  \tau(z q^\perp)=
\tau(z)-\sup_{p\approx s}  \tau(z p)$. Therefore, if $p=s_0$ in
accordance with the constructions of  Lemma \ref{pre1} is chosen
the premises of Proposition \ref{pre} in respect of $\inf_{q\approx
r}  \tau(z q)$ are fulfilled, with $q_0=s_0^\perp$.
Also, since we are in the finite factor case, according to Remark \ref{finfac1}
the hypothesis of Proposition \ref{pre} also applies in the situation at hand, and
then especially shows that formula
($*$) has to be fulfilled. By our preliminaries we then even know that the
asserted formula has to be true also when referring to $x,y\in {\mathcal L}^2(M,\tau)$.
Thus, the assertion holds for a normal
trace on a finite factor.

To see the hypothesis to be true also in the non-factorial cases,
we first remark that owing to normality of $\tau$ also $\tau^x$,
$\tau^y$ and $\tau(|x^*y|(\cdot))$ have to be normal positive
linear forms. It is then obvious from the structure of Definition
\ref{pafi} and definition of $PAIRS_r(M)$ that via central
decomposition techniques $F_M(\tau^x,\tau^y|r)$ can be written as
an appropriately defined central integral with respect to some
central measure $\mu(\lambda)$ over terms
$F_{M_\lambda}(\tau^{x_\lambda},\tau^{y_\lambda}|r_\lambda)$ with
appropriately defined field $\{M_\lambda\}$ of finite `subfactors'
$M_\lambda$ acting over direct integral Hilbert subspaces ${\mathcal H}_\lambda\subset {\mathcal H}$,
with $r=\int \oplus r_\lambda d\/\mu(\lambda)$, $x=\int
\oplus x_\lambda d\/\mu(\lambda)$ and $y=\int \oplus y_\lambda d\/\mu(\lambda)$.
There, the index $\lambda$ refers to the
central decomposition which is isometrically defined relative to some direct integral decomposition
${\mathcal H}=\int\oplus{\mathcal H}_\lambda d\/\mu(\lambda)$ of the Hilbert space of the
$vN$-algebra in question, see e.g.~\cite[{\em Chapter I.},\,5.,\,10.~Corollary]{Schw:67} for the
details. From the factorial case we then have $$
F_{M_\lambda}(\tau^{x_\lambda},\tau^{y_\lambda}|r_\lambda)=
\inf_{q_\lambda\approx r_\lambda} \tau(|x_\lambda^*y_\lambda|
q_\lambda)= \tau(|x_\lambda^*y_\lambda| s_\lambda)\,, $$ for some
orthoprojections $s_\lambda\in M_\lambda$ obeying $s_\lambda\approx
r_\lambda$ and
$\left[s_\lambda,|x_\lambda^*y_\lambda|\right]={\mathbf 0}$, for
all $\lambda$. Hence, in defining $s=\int \oplus s_\lambda
d\/\mu(\lambda)$ gives an orthoprojection $s\in M$ which obeys
$s\approx r$ and $[s,|x^*y|]$, and which minimizes $\tau(|x^*y|q)$
over all $q\approx r$, and thus finally our hypothesis is seen to
hold in each case of a finite $vN$-algebra.
\end{proof}
\subsection*{Partial fidelity on a semifinite $vN$-algebra}
Now the results on partial fidelity will be extended from finite to semifinite $vN$-algebras.
In the factor case, this will be done with the help of Proposition \ref{ap} and its consequences.

\begin{satz}\label{end}
Suppose $M$ is semifinite but properly infinite.
For normal trace $\tau$ and
$x,y\in {\mathcal L}^2(M,\tau)$ the formula
$F_M(\tau^x,\tau^y|r)=\inf_{q\approx r}  \tau(|x^*y|q)$ is valid, for
each orthoprojection $r\in M$.
\end{satz}
\begin{proof}
Note that in view of decomposition theory and by an analogous line of arguments as
used in the proof of
Proposition \ref{finfid},
the nonfactorial cases now can be reduced to the semifinite but properly infinite factorial cases. In line with this,
we may content with giving proofs for semifinite factors of type ${\mathrm I}_\infty$ or
${\mathrm II}_\infty$. Suppose that case for $M$.
To be nontrivial, assume also $\tau\not\equiv 0$. Thus $\tau$ is a faithful normal trace.
First remind that if $r$ is not of finite relative codimension, then
the arguments of the
proof of Lemma \ref{codim} apply and for each $\varepsilon >0$ and positive linear form $\omega$ show
that $s\approx r$ exists with $\omega(s)<\varepsilon$. This especially also applies to
$\omega=\tau(|x^*y|(\cdot))$, and thus in view of the latter and Lemma \ref{codim} the validity of the asserted
formula for $r$ with $\tau(r^\perp)=\infty$ is seen.
Thus, it remains to be seen what happens around this formula, for $r$ with $\tau(r^\perp)<\infty$, that is, for $r$ with finite relative codimension. Also note that the same arguments
as used in the finite case at the beginning of the proof of Proposition \ref{finfid}
can be applied and show that without loss of generality
$x,y\geq {\mathbf 0}$, $x,y\not={\mathbf 0}$ can be supposed in the proof.
In the following we put that case, with $\tau(r^\perp)<\infty$.
Start with showing that under these premises the following estimate is fulfilled:
\begin{equation}\label{gf'}
\inf_{q\approx r}  \tau(|xy|q)\geq F_M(\tau^x,\tau^y|r)\,.
\end{equation}
Let $\varepsilon >0$ be a real. In view of $y\in {\mathcal L}^2(M,\tau)_+$ and by
normality of $\tau$, there has to exist
a finite orthoprojection $Q$ obeying $Qy=yQ$, $\sqrt{\tau^x(Q^\perp)\tau^y(Q^\perp)}\leq \varepsilon$ and $\infty > \tau(Q)\geq \tau(r^\perp)$. Let $P$ be an orthoprojection with $P\leq Q$ and $\tau(P)=\tau(Q)-\tau(r^\perp)$. According to \eqref{cs14} we then have
\begin{subequations}\label{star}
\begin{equation}\label{star0}
F_M(\tau^x,\tau^y)-\varepsilon\leq F_{QMQ}(\tau^x|_{QMQ},\tau^y|_{QMQ}\big|P)\,.
\end{equation}
Note that $\tau^x|_{QMQ}=\tau^{|xQ|}|_{QMQ}$ and
$\tau^y|_{QMQ}=\tau^{|yQ|}|_{QMQ}$ hold. Also, since $QMQ$ is finite and $\tau|_{QMQ}$ is
a normal trace there, in view of Proposition \ref{finfid} we get
$$
F_M(\tau^x,\tau^y|r)\leq \inf_{q\leq Q,\,\tau(q)=\tau(Q)-\tau(r^\perp)} \tau(\bigl||xQ|\,|yQ|\bigr|q)+\varepsilon\,.
$$
Now, by assumption on $y$ and $Q$ one has
$\bigl||xQ|\,|yQ|\bigr|^2=\bigl||xQ|\,yQ\bigr|^2=Qyx^2yQ=Q|xy|^2 Q$.
Thus, the former estimate turns
into the following one\,:
\begin{equation}\label{star2}
F_M(\tau^x,\tau^y|r)\leq
\inf_{q\leq Q,\,\tau(q)=\tau(Q)-\tau(r^\perp)} \tau(\sqrt{Q|xy|^2 Q}\,q)+\varepsilon\,.
\end{equation}
Define $Y=\sqrt{Q|xy|^2 Q}$.
Relating the expression on the r.h.s.~of \eqref{star2}, note that owing to $Q Y=Y$ one
can write as follows\,:
\begin{equation}\label{star3}
\begin{split}
\inf_{q\leq Q,\,\tau(q)=\tau(Q)-\tau(r^\perp)} \tau(Y q)& = \tau(Y Q)-
\sup_{p\leq Q,\,\tau(p) =\tau(r^\perp)} \tau(Y p)\\ & =\tau(Y)-
\sup_{p\leq Q,\,\tau(p)=\tau(r^\perp)} \tau(Y p)\,.
\end{split}
\end{equation}
Note that, for any $p_0$ with $\tau(p_0)=\tau(r^\perp)$, $Qp_0Q\leq {p'}_0\leq Q$ must be fulfilled,
for some ${p'}_0\prec p_0$ (note that $s(Qp_0Q)\sim s(p_0Qp_0)\leq p_0$). Hence, there exists $p_1\leq Q$
with $Qp_0Q\leq p_1$ and  $\tau(p_1)=\tau(p_0)=\tau(r^\perp)$. Thus we may conclude that
\begin{equation}\label{star4}
\sup_{p\leq Q,\,\tau(p)=\tau(r^\perp)} \tau(Y p)=\sup_{\tau(p)=\tau(r^\perp)} \tau(Y QpQ)=
\sup_{\tau(p)=\tau(r^\perp)} \tau( Y p)\,.
\end{equation}
\end{subequations}
But then from the previous and \eqref{star3} we infer
\begin{equation}\label{cs0}
\inf_{q\leq Q,\,\tau(q)=\tau(Q)-\tau(r^\perp)} \tau(Y q)
=\tau(Y)-
\sup_{\tau(p)=\tau(r^\perp)} \tau(Y p)
=\inf_{q\approx r} \tau(Y q)\,.
\end{equation}
Note that $Y=\sqrt{Q|xy|^2 Q}=|\,|xy|Q\,|$ has finite support $s(Y)$. Hence, also
$X=\sqrt{|xy| Q|xy| }=|(|xy|Q)^*|$ has finite
support $s(X)$, with $s(X)\sim s(Y)$. Thus $s(X)\approx s(X)$, and therefore there is unitary $u\in M$
transforming $X$ into $Y$, $Y=u X u^*$. Hence, $\inf_{q\approx r} \tau(Y q)=\inf_{q\approx r} \tau(X q)$
follows. Finally, by operator monotonicity of the square root $X\leq |xy|$ must be fulfilled. Hence
$\inf_{q\approx r} \tau(X q)\leq \inf_{q\approx r} \tau(|xy| q)$, and thus
we end up with the estimate
$$
\inf_{q\leq Q,\,\tau(q)=\tau(Q)-\tau(r^\perp)} \tau(Y q)\leq \inf_{q\approx r} \tau(|xy| q)\,.
$$
From this in view of \eqref{star2} the estimate $F_M(\tau^x,\tau^y|r)\leq
\inf_{q\approx r} \tau(|xy| q)+\varepsilon$ is obtained.
Since $\varepsilon > 0$ can be chosen at will, \eqref{gf'} for $x,y\geq {\mathbf 0}$ is seen to hold.

To see that also
\begin{equation}\label{gf''}
F_M(\tau^x,\tau^y|r)\geq \inf_{q\approx r} \tau(|xy| q)\,,
\end{equation}
holds, for $x,y\geq {\mathbf 0}$ and $r$ with $\tau(r^\perp)<\infty$,
note that in line with Proposition \ref{ap}\,\eqref{ap1} for $\varepsilon>0$ there exists
an orthoprojection $p\approx r$ and an ascending sequence $\{Q_n\}$ of finite orthoprojections
with $\bigvee_n Q_n={\mathbf 1}$ and $p\geq Q_k^\perp$, for each $k\in {\mathbb N}$, such that
\begin{subequations}\label{cs'}
\begin{equation}\label{cs1}
F_M(\omega,\varrho|r)+\varepsilon\geq \limsup_{n\to\infty}
F_{Q_nMQ_n}\bigl(\omega|_{Q_nMQ_n},\varrho|_{Q_nMQ_n}\big|\,p-Q_n^\perp\bigr)
\end{equation}
is fulfilled. Thus, in line with Proposition \ref{finfid}, since owing to $p\approx r$
$\tau(p-Q_n^\perp)=\tau(Q_n)-\tau(p^\perp)=\tau(Q_n)-\tau(r^\perp)$ holds and
$\tau^x|_{Q_nMQ_n}=(\tau|_{Q_nMQ_n})^{|xQ_n|}$ and $\tau^y|_{Q_nMQ_n}=(\tau|_{Q_nMQ_n})^{|yQ_n|}$
are fulfilled, \eqref{cs1} can be rewritten into the following form\,:
\begin{equation}\label{cs3}
F_M(\tau^x,\tau^y|r)+\varepsilon\geq
 \limsup_{n\to\infty}\inf_{q\leq Q_n, \tau(q)=\tau(Q_n)-\tau(r^\perp)}\tau(\bigl||xQ_n|\,|yQ_n|\bigr|q)\,.
\end{equation}
Clearly, owing to finiteness of $Q_n$ and $r^\perp$, the conclusions which led us to see that
\eqref{cs0} holds can be applied
with $Q_n$ and
$\bigl|x|yQ_n|\bigr|=\bigl||xQ_n|\,|yQ_n|\bigr|$ instead of $Q$ and $Y$ as well, and
analogously yield
$$
\inf_{q\leq Q_n, \tau(q)=\tau(Q_n)-\tau(r^\perp)}\tau(\bigl||xQ_n|\,|yQ_n|\bigr|q)=
\inf_{q\approx r} \tau(\bigl||xQ_n|\,|yQ_n|\bigr|q)=\inf_{q\approx r} \tau(\bigl|x|yQ_n|\bigr|q)\,.
$$
Thus, in view of \eqref{cs3},
$
F_M(\tau^x,\tau^y|r)+\varepsilon\geq  \limsup_{n\to\infty}\inf_{q\approx r} \tau(\bigl|x|yQ_n|\bigr|q)$ follows, for $n\geq n(\varepsilon)$.
Now, define $x_m=x Q_m$, for each
$m\in {\mathbb N}$. Then $\tau(s(|x_m|))<\infty$, and $\bigl|x_m^*|yQ_n|\bigr|\leq \bigl|x|yQ_n|\bigr|$, by operator monotony of the square
root. Thus one arrives at
\begin{equation}\label{cs4}
F_M(\tau^x,\tau^y|r)+\varepsilon\geq
 \limsup_{n\to\infty}\inf_{q\approx r} \tau(\bigl|x_m^*|yQ_n|\bigr|q)\,,
\end{equation}
for each $m\in {\mathbb N}$ and $n\geq n(\varepsilon)$.
Now, for each orthoprojection $q$ one has
$$\bigl|\tau(|x_m^* y|q)-\tau(\bigl|x_m^*|yQ_n|\bigr|q)\bigr|\leq \bigl\|\tau(\{|x_m^* y|-
\bigl|x_m^*|yQ_n|\bigr|\}(\cdot))\bigr\|_1\,,$$
and thus upon applying Lemma \ref{confi}\,\eqref{confi2} with $q_n=Q_n$, $a=x_m$ and
$b=y$,
$$
\inf_{q\approx r} \tau(|x_m^* y|q)= \limsup_{n\to\infty}\inf_{q\approx r} \tau(\bigl|x_m^*|yQ_n|\bigr|q)
$$
can be followed.
Hence, in view of \eqref{cs4},
for each $m\in {\mathbb N}$ one infers
\begin{equation}\label{cs5}
F_M(\tau^x,\tau^y|r)+\varepsilon\geq \inf_{q\approx r} \tau(\bigl|x_m^* y\bigr|q)
\,.
\end{equation}
\end{subequations}
Upon applying Lemma \ref{confi}\,\eqref{confi1}, with $q_n=Q_n$, $a=x$, $b=y$ and paying attention
to
$$
\bigl|\tau(|x^*y|q)-\tau(\bigl|x_m^* y\bigr|q)\bigr|\leq \bigl\|\tau(\{|x^*y|-\bigl|x_m^*y\bigr|\}(\cdot))\bigr\|_1
$$
which holds uniformly in the orthoprojection $q$, we see $\lim_{m\to\infty}\inf_{q\approx r} \tau(\bigl|x_m^* y\bigr|q)=\inf_{q\approx r} \tau(|x^* y|q)=\inf_{q\approx r} \tau(|xy|q)$ (remind $x=x^*$).
Hence in view of \eqref{cs5} the estimate
$F_M(\tau^x,\tau^y|r)+\varepsilon\geq \inf_{q\approx r} \tau(|xy|q) $ is obtained.
Since $\varepsilon>0$
could be chosen at will, \eqref{gf''} is seen to hold, in the special
case with $x,y\geq {\mathbf 0}$ and $\tau(r^\perp)<\infty$.
In combining  \eqref{gf'} and  \eqref{gf''} yields that the assertion is true, in these
special situations on a factor of type ${\mathrm I}_\infty$ or ${\mathrm II}_\infty$. In view of our preliminaries the assertion then has to be true also in general, for normal trace $\tau$ on
a semifinite but properly
infinite $vN$-algebra $M$ on a separable Hilbert space ${\mathcal H}$,
any $x,y \in {\mathcal L}^2(M,\tau)$, and any orthoprojection $r\in M$.
\end{proof}
\subsection*{Conclusions}
Note that any $vN$-algebra $M$ uniquely decomposes into the direct sum of
three parts which are finite, properly infinite and semifinite, and
purely infinite, respectively. Since a normal trace is vanishing on the purely infinite
part, in view of \eqref{esti0.2}, Proposition \ref{finfid} and Proposition \ref{end}
one can conclude that Theorem \ref{haupt} is true, that is, the $vN$-algebraic
version of Uhlmann's formula \eqref{finfid.1} holds.

\renewcommand{\appendixpagename}{Appendix}
\renewcommand{\appendixtocname}{Appendix}
\footnotesize
\appendixpage
\begin{appendix}
\setcounter{theorem}{0}
\setcounter{corolla}{0}
\setcounter{remark}{0}
\renewcommand{\thetheorem}{\Alph{theorem}}
\renewcommand{\thecorolla}{\Alph{corolla}}
\renewcommand{\theremark}{\Alph{remark}}
\subsection*{On the notion of fidelity in quantum physics}
Within the usual framework of quantum physics, a quantum channel
can be any quantum device, with input and output states
corresponding to density operators over the Hilbert space
${\mathcal H}$ of the device system. Thereby, depending on the
device, in principle any finite dimensional or separable Hilbert
space ${\mathcal H}$ might occur. Thus, since in the context at
hand usually the algebra of all bounded linear operators ${\mathsf
B}({\mathcal H})$ is considered as algebra of observables, it is
rather reasonable to think of `fidelity' as a functor on objects
$\{{\mathcal H},\varrho_{\text{in}},\varrho_{\text{out}}\}$
consisting of such a Hilbert space ${\mathcal H}$ and an ordered
pair of density operators over this space, and which acts into
${\mathbb R}_+$. To be qualified as a measure of accuracy, it is
reasonable to impose some canonical structural properties on
`fidelity'. Before discussing that matter, in preliminary it is
useful to assemble some majorization-like results for pairs of
density operators.

\subsubsection*{Transforming pairs of density operators by cp-stochastic linear maps}
Let ${\mathcal H}_0$ be an auxiliary infinite dimensional,
separable Hilbert space. Let ${\mathcal U}({\mathcal H}_0)$ be the
group of unitary operators there. On pairs of density operators
over ${\mathcal H}_0$, consider functions
$f:\{\omega,\varrho\}\mapsto f(\omega,\varrho)\in {\mathbb R}_+$
as follows:
\renewcommand{\labelenumii}{\alph{enumi}.\arabic{enumii}}
\renewcommand{\theenumii}{\alph{enumi}.\arabic{enumii}}
\begin{enumerate}
\renewcommand{\labelenumi}{(a.\arabic{enumi})}
\renewcommand{\theenumi}{(a.\arabic{enumi})}
\item\label{add1}
$f$ be uniformly continuous and weakly$^*$--upper semicontinuous;
\item\label{add2}
$f$ be unitarily invariant, i.e.~$f(u\omega u^*, u\varrho u^*)=f(\omega,\varrho)$, for any $u\in {\mathcal U}({\mathcal H}_0)$;
\item\label{add3}
$f$ be (jointly) concave over all pairs of density operators.
\end{enumerate}
Note that to each sequence $\{n_1<n_2<n_3<\cdots\}\subset {\mathbb
N}$ with $n_1|n_2|n_3|\cdots$ ($n_k$ is divisor of $n_{k+1}$, for
each $k$), there exist sequences $\{{\mathcal H}_k\}$,
$\{{\mathcal D}_k\}$ and  $\{{\mathcal H}_k^\prime\}$ of Hilbert
spaces, with $\dim {\mathcal H}_k=n_k$, ${\mathcal H}_0={\mathcal
H}_k\otimes{\mathcal H}_k^\prime$ and ${\mathcal
H}_{k+1}={\mathcal H}_k\otimes {\mathcal D}_k$, for each $k\in
{\mathbb N}$, and such that the subalgebra $\bigcup_k {\mathsf
B}({\mathcal H}_k)\otimes {\mathbf 1}$ is {\em irreducibly} acting
over ${\mathcal H}_0$. Whenever such situation occurs, and $E_k$
is a normal conditional expectation projecting from ${\mathsf
B}({\mathcal H}_0)$ onto ${\mathsf B}({\mathcal H}_k)\otimes
{\mathbf 1}$, in addition the following continuity property will
be required:
\renewcommand{\labelenumi}{(a.\arabic{enumi})}
\renewcommand{\theenumi}{(a.\arabic{enumi})}
\begin{enumerate}
\setcounter{enumi}{3}
\item\label{add6}
$f(\omega,\varrho)=\lim_{k\to\infty} f(\omega\circ E_k,\varrho\circ E_k)\,.$
\end{enumerate}
Trace class operators over a Hilbert space ${\mathcal H}$ will be
denoted by ${\mathcal T}({\mathcal H})$. Let $\omega_k\in
{\mathcal T}({\mathcal H}_k)$, $\sigma_k\in {\mathcal T}({\mathcal
H}_k^\prime)$ refer to the unique density operators obeying $\tr_k
\omega_k A=\tr \omega( A\otimes {\mathbf 1})$, $E_k({\mathbf
1}\otimes B)= \{\tr_k^\prime \sigma_k B\}\,{\mathbf 1}$, for each
$A\in {\mathsf B}({\mathcal H}_k)$ and $B\in {\mathsf B}({\mathcal
H}_k^\prime)$, with the standard traces $\tr_k$, $\tr_k^\prime$
and $\tr$ over ${\mathcal H}_k, \,{\mathcal H}_k^\prime$ and
${\mathcal H}_0$, respectively. Then, whereas
$\{\omega_k,\varrho_k\}$ does not depend on $E_k$, depending from
the latter, $\sigma_k$ can be any density operator over ${\mathcal
H}_k^\prime$. Hence, \ref{add6} is equivalent to the condition
\renewcommand{\labelenumi}{(a.\arabic{enumi}$^\prime$)}
\renewcommand{\theenumi}{(a.\arabic{enumi}$^\prime$)}
\begin{enumerate}
\setcounter{enumi}{3}
\item\label{add7}
$f(\omega,\varrho)=\lim_{k\to\infty} f(\omega_k\otimes\sigma_k,\varrho_k\otimes\sigma_k),\,\forall\,\{\sigma_k\}\,.$
\end{enumerate}
Remarkably, for $f$ obeying \ref{add1}--\ref{add6} in addition the following property holds:
\renewcommand{\labelenumi}{(\alph{enumi})}
\renewcommand{\theenumi}{(\alph{enumi})}
\begin{enumerate}
\setcounter{enumi}{1}
\item\label{b}
$f(\Phi(\omega),\Phi(\varrho)\bigr)\geq f(\omega,\varrho)\,,$
\end{enumerate}
for each {\em completely positive}, {\em trace-preserving} linear
map $\Phi:{\mathcal T}({\mathcal H}_0)\rightarrow {\mathcal
T}({\mathcal H}_0)$. Refer to such mapping as {\em cp-stochastic
linear map}, see
e.g.\,\cite{stine:55,krau:71,koss:72,koss:73,lind:75,lind:76,lind:82}
for notions and use in quantum physics. For a proof of \ref{b},
refer to the first part of the proof of \cite[{\sc
Proposition}]{albe:86.3} (put ${\mathcal M}={\mathcal N}={\mathsf
B}({\mathcal H}_0)$ and $n=2$ there). But more than this is true:
in view of \cite[Theorem
2,\,(2)$\Leftrightarrow$(6),\,(6-1)--(6-9)]{albe:86.2} one infers
that monotonicity of all $f$ with \ref{add1}--\ref{add6} is
characteristic of situations as in \ref{b}.
\renewcommand{\labelenumi}{(\arabic{enumi})}
\renewcommand{\theenumi}{\arabic{enumi}}
\begin{theorem}\label{c}
Under the given premises, the following are mutually
equivalent\textup{:}\normalsize
\begin{enumerate}\footnotesize
\item\label{c1}
$f\bigl(\omega^\prime,\varrho^\prime\bigr)\geq f(\omega,\varrho),\,\forall\,f\text{ obeying \textup{\ref{add1}--\ref{add6}/\ref{add7}}}\,;$
\item\label{c2}
$\omega^\prime=\Phi_0(\omega),\,\varrho^\prime=\Phi_0(\varrho),\text{ for cp-stochastic linear map }\Phi_0:{\mathcal T}({\mathcal H}_0)\rightarrow {\mathcal T}({\mathcal H}_0)\,.$
\end{enumerate}
\setcounter{enumi}{0}
\end{theorem}
It is interesting that along with \eqref{c1} some redundancy comes
in. In fact, for example, we may restrict to all those $f$ which
satisfy the following conditions:
\renewcommand{\labelenumi}{(c.\arabic{enumi})}
\renewcommand{\theenumi}{c.\arabic{enumi}}
\begin{enumerate}
\item\label{d1}
$f$ obeys \textup{\ref{add1}--\ref{add6}}\,;
\item\label{d2}
$0\leq f\leq 1$, with $f(\omega,\varrho)=0$ iff $\omega\perp\varrho$, and $f(\omega,\varrho)=1$ iff
$\omega=\varrho$\,;
\item\label{d3}
$f(\omega,\varrho)=|\langle\varphi,\psi\rangle|$, on {\em pure}
states $\omega,\,\varrho$ corresponding to vectors
$\varphi,\psi\in {\mathcal H}_0$,
\end{enumerate}
with the scalar product $\langle\varphi,\psi\rangle$ of vectors
$\varphi,\psi\in{\mathcal H}_0$, and orthogonality relation
`$\perp$' between states. Also, in context of a Hilbert space
${\mathcal H}$, symbols like
$\omega_\varphi,\varrho_\varphi,\sigma_\varphi\ldots$ in the
following all synonymeously may be used to denote the vector state
over ${\mathsf B}({\mathcal H})$ generated by the unit vector
$\varphi$, or the corresponding pure state density operator,
accordingly.
\begin{corolla}\label{e0}
The following are mutually equivalent\textup{:}\normalsize
\renewcommand{\labelenumi}{(\arabic{enumi})}
\renewcommand{\theenumi}{\arabic{enumi}}
\begin{enumerate}\footnotesize
\item\label{e1}
$f\bigl(\omega^\prime,\varrho^\prime\bigr)\geq f(\omega,\varrho),\,\forall\,f\text{ obeying \textup{\eqref{d1}--\eqref{d3}}}\,;$
\item\label{e2}
$\omega^\prime=\Phi_0(\omega),\,\varrho^\prime=\Phi_0(\varrho),\text{ for cp-stochastic linear map }\Phi_0:{\mathcal T}({\mathcal H}_0)\rightarrow {\mathcal T}({\mathcal H}_0)\,.$
\end{enumerate}
\end{corolla}
\begin{proof}
In view of \cite[Theorem 2,\,(2)$\Leftrightarrow$(6),\,(6-1)--(6-9)]{albe:86.2}, the functions
$P_{\bf{\lambda}}$ (${\bf{\lambda}}$ extends over some set of subscripts) constructed in \cite[\S 4,\,4.6 and \S 5]{albe:85.2}
obey \eqref{d1} and \eqref{d2}, and fulfil
$P_{\bf{\lambda}}(\omega_\varphi,\varrho_\psi)=|\langle\varphi,\psi\rangle|^2$, for each pair of
(normalized) vectors $\varphi,\psi\in {\mathcal H}_0$. From \cite[5.3]{albe:85.2} in addition one knows that
$P_{\bf{\lambda}}(\omega^\prime,\varrho^\prime\bigr)\geq P_{\bf{\lambda}}(\omega,\varrho)$ for all
${\bf{\lambda}}$ implies \eqref{e2} to be fulfilled. Note that since the (scalar) square
root is a monotoneously increasing, continuous
concave function on the unit interval, each of the properties of \ref{add1}-\ref{add6} and \eqref{d2}
holds also for $f=\sqrt{P_{\bf{\lambda}}}$, but in addition now also \eqref{d3}
is fulfilled. In view of the previous and Theorem \ref{c} the result now follows.
\end{proof}
\subsubsection*{The functorial characterization}
The point about functions obeying \eqref{d1}--\eqref{d3} is that they can be put
in one-to-one
correspondence to certain functors $F=F_f$ over all objects $\{{\mathcal H},\omega,\varrho\}$
consisting of
a separable Hilbert space ${\mathcal H}$, and ordered pair $\{\omega,\varrho\}$ of density operators
over ${\mathcal H}$, with the correspondence being defined by
\begin{equation*}\label{stern}
F_f({\mathcal H},\omega,\varrho)=
f\bigl(\omega\circ \pi^{-1}\circ E,\varrho\circ \pi^{-1}\circ E\bigr)\,,
\tag{$\star$}
\end{equation*}
for each pair of density operators on ${\mathcal H}$. Thereby, $\pi$ and $E$ can be
any faithful representation of ${\mathsf B}({\mathcal H})$ over
${\mathcal H}_0$ and normal
conditional expectation $E$ from ${\mathsf B}({\mathcal H}_0)$ onto
$\pi({\mathsf B}({\mathcal H}))$, respectively. Relating notations, on the
right-hand side of \eqref{stern} density operators are identified to
normal positive linear forms, and $\pi^{-1}$ is defined over $\pi({\mathsf B}({\mathcal H}))$ as inverse to $\pi$
action, accordingly. It is easy to see from \eqref{d1}--\eqref{d3} that
functors $F$ defined in accordance with \eqref{stern} have the following properties:
\renewcommand{\labelenumi}{(d.\arabic{enumi})}
\renewcommand{\theenumi}{d.\arabic{enumi}}
\begin{enumerate}
\item\label{ad1}
$F({\mathcal H},\cdot,\cdot)$ is uniformly continuous and weakly$^*$--upper semicontinuous;
\item\label{ad2}
$F({\mathcal H},u\omega u^*, u\varrho u^*)=F({\mathcal H},\omega,\varrho)$, for any $u\in {\mathcal U}({\mathcal H})$;
\item\label{ad3}
$F({\mathcal H},\cdot,\cdot)$ is (jointly) concave over all pairs of density operators on
${\mathcal H}$;
\item\label{ad4}
$0\leq F\leq 1$, with $F({\mathcal H},\omega,\varrho)=0$ iff $\omega\perp\varrho$, and
$F({\mathcal H},\omega,\varrho)=1$ iff
$\omega=\varrho$\,;
\item\label{ad5}
$F({\mathcal H},\omega_\varphi,\varrho_\psi)=|\langle\varphi,\psi\rangle|$, for
unit vectors $\varphi,\psi\in {\mathcal H}$.
\end{enumerate}
Moreover, in case of $\dim {\mathcal H}=\infty$, and for each
ascendingly directed sequence ${\mathcal H}_1\subset {\mathcal H}_2\subset{\mathcal H}_3\subset \cdots$
of finite dimensional Hilbert spaces, with respect to which a factorization
${\mathcal H}={\mathcal H}_k\otimes {\mathcal H}_k^\prime$ with ${\mathsf B}({\mathcal H}_1)\otimes {\mathbf 1}\subset {\mathsf B}({\mathcal H}_2)\otimes {\mathbf 1}\subset {\mathsf B}({\mathcal H}_3)\otimes {\mathbf 1}\subset\cdots \subset {\mathsf B}({\mathcal H})$ and irreducibly acting $\bigcup_k {\mathsf B}({\mathcal H}_k)\otimes {\mathbf 1}$ over ${\mathcal H}$ exists, in addition
\begin{enumerate}
\setcounter{enumi}{5}
\item\label{ad6}
$F({\mathcal H},\omega,\varrho)=\lim_{k\to\infty} F({\mathcal H}_k,\omega_k,\varrho_k)$
\end{enumerate}
is fulfilled, with the reduced to ${\mathsf B}({\mathcal H}_k)\otimes {\mathbf 1}$ density operators
$\omega_k$ and $\varrho_k$ of $\omega$ and $\varrho$, respectively,
for each $k\in {\mathbb N}$.

On the other hand, yet easier to see, for each functor $F$ obeying
\eqref{ad1}--\eqref{ad6}, in defining $f=f_F$ at each pair $\{\omega,\varrho\}$ of density operators over
${\mathcal H}_0$ by
\begin{equation*}\label{2stern}
f_F(\omega,\varrho)=F({\mathcal H}_0,\omega,\varrho),
\tag{$\star\star$}
\end{equation*}
a function obeying \eqref{d1}--\eqref{d3} arises. Hence, in view of
\eqref{stern} and \eqref{2stern}, those functors (resp.~functions)
obtained by formula \eqref{stern} (resp.~formula \eqref{2stern}) from
functions $f$ (resp.~functors $F$) obeying \eqref{d1}--\eqref{d3} (resp.~\eqref{ad1}--\eqref{ad6})
can be {\em equivalently} characterized as set
of {\em all} functors (resp.~{\em all} functions) over the above
mentioned category of objects (resp.~pairs of
density operators on ${\mathcal H}_0$) and satisfying \eqref{ad1}--\eqref{ad6} (resp.~\eqref{d1}--\eqref{d3}). For all that follows, it is useful to
introduce another notation, in context of two pairs of density operators
$\{\omega,\varrho\}$ and $\{\omega^\prime,\varrho^\prime\}$ over
separable Hilbert spaces ${\mathcal H}$ and ${\mathcal H}^\prime$, respectively. Namely,
whenever a {\em cp}-stochastic linear map $\Phi:{\mathcal T}({\mathcal H})\rightarrow
{\mathcal T}({\mathcal H}^\prime)$ with $\omega^\prime=\Phi(\omega)$ and
$\varrho^\prime=\Phi(\varrho)$ exists, whatever the transforming $\Phi$ be,
then henceforth $\{{\mathcal H}^\prime,\omega^\prime,\varrho^\prime\}\ll_*\{{\mathcal H},\omega,\varrho\}$ will be used to indicate that special situation. A main result about the
functors at hand then reads as follows:
\renewcommand{\labelenumi}{(\arabic{enumi})}
\renewcommand{\theenumi}{\arabic{enumi}}
\begin{theorem}\label{g}
Let ${\mathcal H}$ and ${\mathcal H}^\prime$ be separable Hilbert
spaces, with pairs of density operators $\{\omega,\varrho\}$ and
$\{\omega^\prime,\varrho^\prime\}$, respectively. The following
are mutually equivalent\textup{:}\normalsize
\begin{enumerate}\footnotesize
\item\label{g1}
$F\bigl({\mathcal H}^\prime,\omega^\prime,\varrho^\prime\bigr)\geq F({\mathcal H},\omega,\varrho),\,\forall\,F\text{ obeying \textup{\eqref{ad1}--\eqref{ad6}}}\,;$
\item\label{g2}
$\ \,\{{\mathcal H}^\prime,\omega^\prime,\varrho^\prime\}\ll_*\{{\mathcal H},\omega,\varrho\}\,;$
\end{enumerate}
\end{theorem}
\begin{proof}
Note that owing to the just discussed, condition \eqref{g1} must be equivalent to
$$f\bigl(\omega^\prime\circ {\pi^\prime}^{-1}\circ E^\prime,\varrho^\prime\circ {\pi^\prime}^{-1}\circ E^\prime\bigr)\leq f(\omega\circ \pi^{-1}\circ E,\varrho\circ \pi^{-1}\circ E)\,,$$
for each $f$ obeying \eqref{d1}--\eqref{d3}, with faithful representations $\pi,\pi^\prime$
of ${\mathsf B}({\mathcal H})$, ${\mathsf B}({\mathcal H}^\prime)$ over ${\mathcal H}_0$, and normal
conditional expectations $E,E^\prime$ from ${\mathsf B}({\mathcal H}_0)$ onto $\pi({\mathsf B}({\mathcal H}))$, $\pi^\prime({\mathsf B}({\mathcal H}^\prime))$, respectively.
By Corollary \ref{e0}, the latter estimates are equivalent to the
existence of a {\em cp}-stochastic linear mapping
$\Phi_0$ on
${\mathcal T}({\mathcal H}_0)$ obeying $\omega^\prime\circ {\pi^\prime}^{-1}\circ E^\prime=\Phi_0\bigl(
\omega\circ \pi^{-1}\circ E\bigr)$ and $\varrho^\prime\circ {\pi^\prime}^{-1}\circ E^\prime=\Phi_0\bigl(
\varrho\circ \pi^{-1}\circ E\bigr)$.
Since faithful representations and normal conditional expectations are special unity preserving,
normal completely positive
linear mappings, in defining $\Phi$ by $\Phi(\sigma)=\Phi_0\bigl(\sigma \circ \pi^{-1}\circ E\bigr)\circ {\pi^\prime}$, for each $\sigma\in {\mathcal T}({\mathcal H})$, a normal
{\em cp}-stochastic linear mapping from ${\mathcal T}({\mathcal H})$ into
${\mathcal T}({\mathcal H}^\prime)$ will be given. It is obvious by construction
that $\Phi$ then will do the job of \eqref{g2}.

On the other hand, for each {\em cp}-stochastic linear mapping $\Phi:{\mathcal T}({\mathcal H})\rightarrow {\mathcal T}({\mathcal H}^\prime)$, and given faithful representations $\pi,\pi^\prime$
of ${\mathsf B}({\mathcal H})$, ${\mathsf B}({\mathcal H}^\prime)$ over ${\mathcal H}_0$, and normal
conditional expectations $E,E^\prime$ from ${\mathsf B}({\mathcal H}_0)$ onto $\pi({\mathsf B}({\mathcal H}))$, $\pi^\prime({\mathsf B}({\mathcal H}^\prime))$, through setting
$\Phi_0(\sigma)=\Phi(\sigma\circ \pi)\circ {\pi^\prime}^{-1}\circ E^\prime$, for $\sigma\in {\mathcal T}({\mathcal H}_0)$, a {\em cp}-stochastic linear map
$\Phi_0$ over ${\mathcal T}({\mathcal H}_0)$ is obtained. Thus $\Phi_0(\mu\circ {\pi}^{-1}\circ E)=\Phi(\mu)\circ {\pi^\prime}^{-1}\circ E^\prime$ follows, for each $\mu\in {\mathcal T}({\mathcal H})$. Hence,
by \eqref{stern} and Corollary \ref{e0}, for density operators $\{\omega,\varrho\}$
over ${\mathcal H}$ we infer
$
F_f({\mathcal H}^\prime,\Phi(\omega),\Phi(\varrho))=f(\Phi(\omega)\circ {\pi^\prime}^{-1}\circ E^\prime,\Phi(\varrho)\circ {\pi^\prime}^{-1}\circ E^\prime)=f(\Phi_0(\omega\circ {\pi}^{-1}\circ E),\Phi_0(\varrho\circ {\pi}^{-1}\circ E))\geq f(\omega\circ {\pi}^{-1}\circ E,\varrho\circ {\pi}^{-1}\circ E)=F_f({\mathcal H},\omega,\varrho)$. Since $f$ can be any function obeying \eqref{d1}--\eqref{d3}, according to
the above $F_f$ can be any functor obeying \eqref{ad1}--\eqref{ad6}, and therefore also
\eqref{g2} implies \eqref{g1}.
\end{proof}
Note that \eqref{ad1}--\eqref{ad6} make sense and extend to
properties of functors on larger subcategories of (injective)
$vN$-algebras and their normal states. We then have
$F_M(\omega,\varrho)$ rather than $F({\mathcal
H},\omega,\varrho)$, with triple $\{M,\omega,\varrho\}$ instead of
$\{{\mathcal H},\omega,\varrho\}$, with injective $vN$-algebra $M$
and pair of normal states on $M$. In particular, Theorem \ref{g}
\eqref{g2}$\Rightarrow$\eqref{g1} extends to all {\em injective}
$vN$-algebras; for details see \cite{albe:86.3} and
\cite[3.1]{Albe:92.3}. The best known of all functors obeying
\eqref{ad1}--\eqref{ad6} under rather general conditions is
$F_M=\sqrt{P_M}$, with Uhlmann's $^*$-algebraic transition
probability $P_M$. For {\em atomic} $vN$-algebras, also Theorem
\ref{g} \eqref{g1}$\Rightarrow$\eqref{g2} remains true, and thus
Theorem \ref{g} extends to the category of {\em atomic}
$vN$-algebras at least, see \cite[\S 5.\,Theorem,\,\S
10.\,Bemerkung]{albe:85.3}. The equivalence mentioned on in
context of \eqref{stern}/\eqref{2stern} can be also seen as
special case of \cite[Lemmata 1/2]{albe:85.3} and thus remains
true in this slightly more general context, too.  We do not go
into the details of that extension, but instead refer to
\cite{DeC:71,storm:72,storm:74,albe:85.3}.
\subsubsection*{The minimal functor and related facts--fidelity \`{a} la Jozsa}
Quite remarkably, within all functors obeying \eqref{ad1}--\eqref{ad6} a least element $F_{\min}$
exists \cite[\S 12,\,eqs.\,(21)]{albe:85.3}. That is, for each at most separable Hilbert
space ${\mathcal H}$ and pair of density operators
$\{\omega,\varrho\}$ over ${\mathcal H}$, by
\renewcommand{\labelenumi}{(\alph{enumi})}
\renewcommand{\theenumi}{\alph{enumi}}
\begin{enumerate}
\setcounter{enumi}{4}
\item\label{e}
$F_{\min}({\mathcal H},\omega,\varrho)=\inf\bigl\{ F({\mathcal H},\omega,\varrho):F\text{ with \eqref{ad1}--\eqref{ad6}}\bigr\}$
\end{enumerate}
another functor obeying \eqref{ad1}--\eqref{ad6} is defined. This functor had been identified
as $F_{\min}=\sqrt{P}$,
with the functor $P$ of the {\em generalized transition probability for density operators} invented by
Uhlmann, see \cite{Uhlm:76,AlUh:84,Uhlm:85}, and therefore reads as
\begin{enumerate}
\setcounter{enumi}{5}
\item\label{f}
$F_{\min}({\mathcal H},\omega,\varrho)=\tr \{\varrho^{1/2}\omega\varrho^{1/2}\}^{1/2}=\tr \bigl|\omega^{1/2}\varrho^{1/2}\bigr|\,,$
\end{enumerate}
which at $\{\omega,\varrho_\psi\}$ takes a form, which is useful to keep in mind,
for later use:
\begin{enumerate}
\setcounter{enumi}{6}
\item\label{gg}
$F_{\min}({\mathcal H},\omega,\varrho_\psi)=\sqrt{\langle \omega\psi,\psi\rangle}\,.$
\end{enumerate}
For instance, $F_{\min}$ is related to the {\em Bures distance}
$d_{\mathrm{B}}$ \cite{Bure:69,Uhlm:76} through the formula
\begin{enumerate}
\setcounter{enumi}{7}
\item\label{h}
$d_{\mathrm{B}}({\mathcal H},\omega,\varrho)=\sqrt{2}\bigl\{1-F_{\min}({\mathcal H},\omega,\varrho)\bigr\}^{1/2}\,,$
\end{enumerate}
and which according to \eqref{gg} between pure states
$\omega_\psi$ and $\varrho_\varphi$ reduces to the natural
distance $\pmb|${\boldmath{$\psi$}}-{\boldmath{$\varphi$}}$\pmb|$
\begin{enumerate}
\setcounter{enumi}{8}
\item\label{i}
$d_{\mathrm{B}}({\mathcal H},\omega_\psi,\varrho_\varphi)=
\inf_{\alpha\in{\mathbb R}}\|{\mathrm e}^{{\mathsf
i}\alpha}\psi-\varphi\|=$
$\pmb|${\boldmath{$\psi$}}-{\boldmath{$\varphi$}}$\pmb|$\,,
\end{enumerate}
between the rays or fibres {\boldmath{$\psi$}}$=\{\lambda\psi:
|\lambda|=1\}$ and {\boldmath{$\varphi$}}$=\{\lambda\varphi:
|\lambda|=1\}$, accordingly, see \cite{Barg:64}. 
\begin{remark}
Note that it is not 
only in this particular case ($\rm{U}(1)$-bundle) that the Bures distance is playing 
such an exposed r\^{o}le as a natural distance: it is known also in 
the ${C^*}$-algebraic case that the Bures distance can be interpreted as  
the natural distance between the fibres of implementing vectors 
of states (or even positive linear forms, more generally). For an overview on the underlying ideas 
and generalities up to the ${C^*}$-algebraic case, see \cite{Uhlm:91.2,Uhlm:92.1,Uhlm:93,Uhlm:95,Uhlm:96}, 
for an comprehensive account on the functional analytic methods around the Bures distance as 
fibre distance, see \cite{AlPe:00.2}, 
and for the latest tricky details, examples and counterexamples in context of Bures geometry,  
especially for density operators, see \cite{Pelt:00.1}. 
\end{remark}
There exists
another but equivalent to \eqref{e} characterization of
$F_{\min}$. Thereby, within the formulation of the result,
${\mathcal H}$ and ${\mathcal H}^\prime$ may be any two separable
Hilbert spaces, and $\{\omega_\varphi,\varrho_\psi\}$, for
$\varphi,\psi\in{\mathcal H}$, and
$\{\omega^\prime,\varrho^\prime\}\subset {\mathcal T}({\mathcal
H}^\prime)$, may be pairs of vector states and density operators
over the respective Hilbert spaces, accordingly.
\renewcommand{\labelenumi}{(\arabic{enumi})}
\renewcommand{\theenumi}{\arabic{enumi}}
\begin{theorem}\label{rein}
$F=F_{\min}$ is the unique functor obeying
\eqref{ad1}--\eqref{ad6} and such that the following statements
are mutually equivalent\textup{:} \normalsize
\begin{enumerate}\footnotesize
\item\label{rein1}
$\ \ \{{\mathcal H}^\prime,\omega^\prime,\varrho^\prime\}\ll_*
\{{\mathcal H},\omega_\varphi,\varrho_\psi\}$;
\item\label{rein2}
$\ F({\mathcal H}^\prime,\omega^\prime,\varrho^\prime)\geq
F({\mathcal H},\omega_\varphi,\varrho_\psi)$;
\item\label{rein3}
$d_{\mathrm{B}}({\mathcal H}^\prime,\omega^\prime,\varrho^\prime)\leq
d_{\mathrm{B}}({\mathcal H},\omega_\varphi,\varrho_\psi)$.
\end{enumerate}
\end{theorem}
\begin{proof}
Note that owing to \eqref{h}, \eqref{rein3} is equivalent to \eqref{rein2} in the special case of
$F=F_{\min}$. Thus, in view of Theorem \ref{g} and formula \eqref{h}, only
\eqref{rein2}$\Rightarrow$\eqref{rein1} is in quest.
Assume $F\not=F_{\min}$. Then, owing to the common boundary condition \eqref{ad5} at vector states, a separable Hilbert space ${\mathcal H}^\prime$ with
$\dim {\mathcal H}^\prime>1$ and pair of
density operators $\{\omega^\prime,\varrho^\prime\}$ obeying $F({\mathcal H}^\prime,\omega^\prime,\varrho^\prime)>F_{\min}({\mathcal H}^\prime,\omega^\prime,\varrho^\prime)$ have to exist. Thereby,
in view of \eqref{ad4} and by
uniform continuity \eqref{ad1} of $F$ and $F_{\min}$,
$\{\omega^\prime,\varrho^\prime\}$ may be even supposed with $1>F({\mathcal H}^\prime,\omega^\prime,\varrho^\prime)>F_{\min}({\mathcal H}^\prime,\omega^\prime,\varrho^\prime)>0$. Especially, by
$0\leq F\leq 1$ there have to exist unit vectors $\psi,\varphi\in {\mathcal H}^\prime$ such that $F({\mathcal H}^\prime,\omega^\prime,\varrho^\prime)=|\langle\psi,\varphi\rangle|$. Hence, in defining
$\{\omega,\varrho\}=\{\omega_\varphi,\varrho_\psi\}$ we have $F({\mathcal H}^\prime,\omega^\prime,\varrho^\prime)\geq F({\mathcal H}^\prime,\omega_\varphi,\varrho_\psi)$. Assume there were a  {\em cp}-stochastic normal
linear map $\Phi$ over ${\mathcal T}({\mathcal H}^\prime)$, with $\omega^\prime=\Phi(\omega_\varphi)$ and
$\varrho^\prime=\Phi(\varrho_\psi)$, simultaneously.
We then had
$F_{\min}({\mathcal H}^\prime,\omega^\prime,\varrho^\prime)\geq F_{\min}({\mathcal H}^\prime,\omega_\varphi,\varrho_\psi)
= |\langle\psi,\varphi\rangle|$, by Theorem \ref{g} and \eqref{ad5}. This contradicts the
choice of $\varphi,\psi\in {\mathcal H}^\prime$.
Hence, in case of $F\not=F_{\min}$, the implication \eqref{rein2}$\Rightarrow$\eqref{rein1}
cannot be true generally.
On the other hand, in view of Corollary \ref{e0}
and owing to the bijection established by
\eqref{stern} and \eqref{2stern}, the implication in question for some $F$ is equivalent to the
same implication for $f_F$ in the special case with ${\mathcal H}_0={\mathcal H}={\mathcal H}^\prime$.
For $F=F_{\min}$ the latter implication is true (and so is also \eqref{rein3}$\Rightarrow$\eqref{rein1}),
by \cite[Lemma 3]{Albe:82} (see also
\cite[{\sc Theorem 5}]{AlUh:83}).
\end{proof}
\subsubsection*{Fidelity \`{a} la Jozsa and channels with mixed output states}
Note that according to Theorem \ref{rein}, the functor $F_{\min}$, i.e.~fidelity \`{a} la Jozsa, 
is of paramount importance in context of $\ll_*$. In fact, more than the mere equivalence
\renewcommand{\labelenumi}{(\alph{enumi})}
\renewcommand{\theenumi}{\alph{enumi}}
\begin{enumerate}
\setcounter{enumi}{9}
\item\label{j}
$
F_{\min}({\mathcal H}^\prime,\omega^\prime,\varrho^\prime)\geq F_{\min}({\mathcal H},\omega_\varphi,\varrho_\psi)\Longleftrightarrow
\{{\mathcal H}^\prime,\omega^\prime,\varrho^\prime\}\ll_*\{{\mathcal H},\omega_\varphi,\varrho_\psi\}$
\end{enumerate}
is established, by Theorem \ref{rein}\,; at the same stroke it is telling
that implications like
\begin{equation*}\label{3star}
F({\mathcal H}^\prime,\omega^\prime,\varrho^\prime)\geq F({\mathcal H},\omega,\varrho)\,\Longrightarrow\,\{{\mathcal H}^\prime,\omega^\prime,\varrho^\prime\}
\ll_*\{{\mathcal H},\omega,\varrho\}
\tag{$\triangle$}
\end{equation*}
for $F\not=F_{\min}$ cannot hold, in general. Now, we are going to discuss what
happens for $F=F_{\min}$ in \eqref{3star}, also in situations beyond of \eqref{j}. 
That is, we ask whether 
the exclusive use of fidelity \`{a} la Jozsa could be sacrified also 
for the sake of comparing two quantum channels, other than those which are covered by \eqref{j}, 
in the more general situation with both output states being mixed ones. In particular,
the behavior of the relation $\ll_*$ will be studied in restriction to
the generic subclasses ${\mathcal P}_0$ and ${\mathcal P}$ of all objects of the form 
$\{{\mathcal H},\omega_\varphi,\varrho_\psi\}$ and
$\{{\mathcal H},\omega,\varrho_\psi\}$, respectively. It should be emphasized that these
subclasses are the ones
which are known to be of paramount
importance for use with quantum channels, in context of quantum information processing. 
Especially, the class ${\mathcal P}$ is of main interest since it covers the more realistic 
case of the quantum channel with pure input states and admiting also 
mixed output states. Clearly, when considered in restriction 
to ${\mathcal P}_0$, the implication within \eqref{3star}
has to be true, according to Theorem \ref{rein} and since $F=F_{\min}$ holds over ${\mathcal P}_0$. 
On a first view things look hopefully also over ${\mathcal P}$, since we have the following result.
\begin{corolla}\label{nurJoz}
For any fidelity functor $F$ obeying \eqref{ad1}-\eqref{ad6} and differing from $F_{\min}$ 
on ${\mathcal P}$ the implication \eqref{3star} in restriction to ${\mathcal P}$ fails to be true. 
\end{corolla} 
\begin{proof}
Suppose the implication $F({\mathcal H}^\prime,\omega^\prime,\varrho_{\psi'}^\prime)\geq
F({\mathcal H},\omega,\varrho_\psi)\ \Longrightarrow\ \{{\mathcal H}^\prime,\omega^\prime,\varrho_{\psi'}^\prime\}\ll_*
\{{\mathcal H},\omega,\varrho_\psi\}$ were generally true, on ${\mathcal P}$, for some $F$. Then, in the 
special cases with $\omega = \omega_\varphi$, from Theorems \ref{rein}/\ref{g} the 
following equivalence had to be inferred to hold: 
$$F({\mathcal H}^\prime,\omega^\prime,\varrho_{\psi'}^\prime)\geq
F({\mathcal H},\omega_\varphi,\varrho_\psi)\ \Longleftrightarrow\ F_{\min}({\mathcal H}^\prime,\omega^\prime,\varrho_{\psi'}^\prime)\geq
F_{\min}({\mathcal H},\omega_\varphi,\varrho_\psi)\,.$$
In accordance with the premises on $F$, let 
$\{{\mathcal H}^\prime,\omega^\prime,\varrho_{\psi'}^\prime\}$ be chosen such that 
$F({\mathcal H}^\prime,\omega^\prime,\varrho_{\psi'}^\prime)\not=F_{\min}({\mathcal H}^\prime,\omega^\prime,\varrho_{\psi'}^\prime)$. By \eqref{e} we then even know that 
\begin{equation*}\label{4star}
F({\mathcal H}^\prime,\omega^\prime,\varrho_{\psi'}^\prime) > F_{\min}({\mathcal H}^\prime,\omega^\prime,\varrho_{\psi'}^\prime)\,.
\tag{$\triangle\triangle$}
\end{equation*}
Now, let ${\mathcal H}$ be any (at least two-dimensional) Hilbert space, and unit vectors $\varphi,\psi\in 
{\mathcal H}$ chosen as to obey $F({\mathcal H}^\prime,\omega^\prime,\varrho_{\psi'}^\prime)=|\langle\varphi,\psi\rangle|$. According to \eqref{ad4} this choice can always be achieved. But then, owing to  
$\{{\mathcal H},\omega_\varphi,\varrho_\psi\}\in {\mathcal P}_0$, the previous equality also amounts to 
$F({\mathcal H}^\prime,\omega^\prime,\varrho_{\psi'}^\prime)=
F_{\min}({\mathcal H},\omega_\varphi,\varrho_{\psi})=F({\mathcal H},\omega_\varphi,\varrho_{\psi})$. 
Hence, in view of the above equivalence, $F_{\min}({\mathcal H}^\prime,\omega^\prime,\varrho_{\psi'}^\prime)\geq F_{\min}({\mathcal H},\omega_\varphi,\varrho_{\psi})=F({\mathcal H}^\prime,\omega^\prime,\varrho_{\psi'}^\prime)$ had to be followed. This obviously contradicts \eqref{4star}. 
\end{proof}
Finally, we are going to show that the implication within \eqref{3star} even for Jozsa's fidelity 
$F=F_{\min}$ cannot turn true, on the whole class ${\mathcal P}$. After having done so, in view of 
Corollary \ref{nurJoz} we will 
know that in fact there cannot exist any single functor $F$ satisfying  \eqref{ad1}--\eqref{ad6} 
and turning implication \eqref{3star} into a truth, on the whole of ${\mathcal P}$. 
We are going to prove a separation result, for density operators
$\omega\in {\mathcal T}({\mathcal H})$ and $\omega^\prime\in
{\mathcal T}({\mathcal H}^\prime)$, over separable Hilbert spaces ${\mathcal H}$ and
${\mathcal H}^\prime$. Relating notations, ${\mathop{\mathrm{spec}}}(x)$ and ${\mathcal W}(x)$
will be used for
`spectrum' and `numerical range' of $x\in {\mathsf B}({\mathcal H})$,
respectively. Remind
${\mathcal W}(x)=\{\langle x\varphi,\varphi\rangle:\varphi\in {\mathcal H},\,\|\varphi\|=1\}$,
and that $\overline{{\mathcal W}(x)}=\ccon{{\mathop{\mathrm{spec}}}(x)}$ (closed convex hull)
is fulfilled, for normal $x$. In particular, for hermitian $x$,
${\mathcal W}(x)\subset {\mathbb R}$
will be an interval. Let ${\mathcal W}(x)^0$ then be the open interval of the interior points.
\begin{corolla}\label{kanal}
Suppose ${\mathcal W}(\omega^\prime)^0\cap{\mathop{\mathrm{spec}}}(\omega)\not=\emptyset$.
Then, unit vectors
$\varphi\in {\mathcal H}^\prime$ and $\psi\in {\mathcal H}$ with
$F_{\min}({\mathcal H}^\prime,\omega^\prime,\varrho_\varphi^\prime)\geq
F_{\min}({\mathcal H},\omega,\varrho_\psi)$ but
$\{{\mathcal H}^\prime,\omega^\prime,\varrho_\varphi^\prime\}\not\ll_*
\{{\mathcal H},\omega,\varrho_\psi\}$ exist.
\end{corolla}
\begin{proof}
There are orthonormal systems $\{\psi_j\}$ and $\{\varphi_k\}$ within
${\mathcal H}$ and  ${\mathcal H}^\prime$, respectively, such that
$\omega=\sum_j \lambda_j \omega_{\psi_j}$ and  $\omega^\prime=\sum_k \lambda_k^\prime
\omega_{\varphi_k}^\prime$, with $\lambda_1\geq \lambda_2\geq \ldots > 0$ and
$\lambda_1^\prime\geq \lambda_2^\prime\geq \ldots > 0$, $\sum_j \lambda_j=\sum_k \lambda_k^\prime=1$.
The relations between spectrum and numerical range yield
${\mathcal W}(\omega^\prime)^0=\bigl]\lambda_{\inf}^\prime,\lambda_1^\prime\bigr[$, with
$\lambda_{\inf}^\prime=\inf \{\lambda\in {\mathop{\mathrm{spec}}}(\omega^\prime)\}$, and
the supposition ${\mathcal W}(\omega^\prime)^0\cap{\mathop{\mathrm{spec}}}(\omega)\not=\emptyset$
is telling that subscripts $j,k$ with
$\lambda_1^\prime> \lambda_j>\lambda_k^\prime$ have to exist (thus, in particular, both density operators
have to be mixed ones). Thus $\lambda_j\in \bigl]\lambda_k^\prime,\lambda_1^\prime\bigr[$, and
there has to be unique $\beta\in \bigl]0,1\bigr[$ obeying
$\lambda_j=\beta \lambda_1^\prime+(1-\beta)\lambda_k^\prime$.
Note that, according to formula \eqref{gg}, for unit vectors
$\varphi\in {\mathcal H}^\prime,\,\psi\in {\mathcal H}$, one has $F_{\min}({\mathcal H},\omega,\varrho_\psi)^2=
\sum_j \lambda_j \beta_j(\psi),\,F_{\min}({\mathcal H}^\prime,\omega^\prime,\varrho_\varphi^\prime)^2=
\sum_k \lambda_k^\prime \beta_k^\prime(\varphi)$, with non-negative reals
$\beta_j=\beta_j(\psi)=|\langle\psi,\psi_j\rangle|^2,\,\beta_k^\prime=\beta_k^\prime(\varphi)=|\langle\varphi,\varphi_k\rangle|^2$ obeying
$\sum_j \beta_j=\sum_k \beta_k^\prime=1$. We define special unit vectors by setting
$\varphi=\sqrt{\beta}\,\varphi_1+\sqrt{1-\beta}\,\varphi_k$ and $\psi=\psi_j$, and then are in the
case with $\beta_l(\psi)=\beta\,\delta_{jl}$ and
$\beta_r^\prime(\varphi)=\beta\,\delta_{r1}+(1-\beta)\,\delta_{rk}$.
With the help of the previous formulae we infer that
$F_{\min}({\mathcal H},\omega,\varrho_{\psi})=\sqrt{\lambda_j}=
F_{\min}({\mathcal H}^\prime,\omega^\prime,\varrho_\varphi^\prime)$ holds. Also, by construction
$(\omega-\lambda_j\varrho_\psi)\geq 0$.  We are going to show that
$(\omega^\prime-\lambda_j\varrho_\varphi^\prime)$ cannot
be positive. As usually, in the following normal linear forms will be identified with
operators of trace class (thus, $\varrho_\psi$ and $\varrho_\varphi^\prime$
correspond to one-dimensional orthoprojections).
In this sense, note first that $(\omega^\prime-\lambda_j\varrho_\varphi^\prime)=a+b$,
with hermitian trace class operators
$a=(\lambda_1^\prime\omega_{\varphi_1}^\prime+\lambda_k^\prime\omega_{\varphi_k}^\prime
-\lambda_j\varrho_\varphi^\prime)$ and $b=\sum_{r\not=1,k}\lambda_r^\prime \omega_{\varphi_r}^\prime$.
Thereby, $b\geq {\mathbf 0}$ holds, with range of $b$ being orthogonal to the range of $a$. In fact,
owing to $\varphi\in \{\omega_{\varphi_1}^\prime+ \omega_{\varphi_k}^\prime\}{\mathcal H}^\prime$
the hermitian operator
$a=(\lambda_1^\prime\omega_{\varphi_1}^\prime+\lambda_k^\prime\omega_{\varphi_k}^\prime
-\lambda_j\varrho_\varphi^\prime)$ can be of rank two at most, with range in
$\{\omega_{\varphi_1}^\prime+ \omega_{\varphi_k}^\prime\}{\mathcal H}^\prime$.
Hence, in order to prove that $(\omega^\prime-\lambda_j\varrho_\varphi^\prime)$ cannot
be positive, it suffices to prove that $a$ cannot be positive.
This will be done by demonstrating that assuming $a\geq {\mathbf 0}$ will lead into contradictions.
In fact, by construction
$\langle a\varphi,\varphi\rangle=0$. Hence, on assuming positivity of $a$, in view of the
above facts about $a$ and $\varphi$,  as a positive operator $a$ had to be of rank one, with
$a \varrho_\varphi^\prime={\mathbf 0}$,
and thus $a$ besides the defining expression would also have to obey
$a=(\lambda_1^\prime+\lambda_k^\prime -\lambda_j)(\omega_{\varphi_1}^\prime+ \omega_{\varphi_k}^\prime-
\varrho_\varphi^\prime)$. But both
expressions for $a$ cannot hold at the same time, for this contradicts to the obvious fact that
$(\lambda_k^\prime-\lambda_j)\omega_{\varphi_1}^\prime+(\lambda_1^\prime-\lambda_j)\omega_{\varphi_k}^\prime\not=
(\lambda_1^\prime+\lambda_k^\prime)\varrho_\varphi^\prime$.
Thus, $a$ cannot be positive, and therefore by the above also $a+b$ cannot be positive. Hence
$\Phi(\omega-\lambda_j\varrho_\psi)\not=a+b=\omega^\prime-\lambda_j\varrho_\varphi^\prime$, for any
stochastic linear map $\Phi$ (which all are positivity preserving maps). Thus, in particular also
$\{{\mathcal H}^\prime,\omega^\prime,\varrho_\varphi^\prime\}\not\ll_*
\{{\mathcal H},\omega,\varrho_\psi\}$, which closes the proof (see also \cite{AlUh:80} for
a similar proof in case of $2\times 2$-matrices).
\end{proof}
Note that according to formulae \eqref{f}/\eqref{gg} the fundamental relation
$$\bigl\{F_{\min}({\mathcal H},\omega,
\varrho):\forall\,\varrho\bigr\}=\bigl\{F_{\min}({\mathcal H},\omega,
\varrho_\psi):\psi\in {\mathcal H},\|\psi\|=1\bigr\}=\bigl\{\sqrt{\lambda}:\lambda\in
{\mathcal W}(\omega)\bigr\}$$ is fulfilled in respect of the numerical range of $\omega$,
for the range of $F_{\min}$ whenever $\omega$ is held fixed on
some Hilbert space ${\mathcal H}$.
Also, since for faithful density operators
$\omega$ and $\omega^\prime$ on infinite dimensional Hilbert spaces, in the notations of the proof
of Corollary \ref{kanal}, ${\mathcal W}(\omega^\prime)^0=\bigl]0,x_1^\prime\bigr[$ and
$0=\inf_j x_j$ hold, under these conditions for instance the premises of Corollary \ref{kanal}
are always satiesfied. Therefore,
the assertion of Corollary \ref{kanal} does not only show that a functor $F$ obeying
\eqref{d1}--\eqref{d3} with
$F\not=F_{\min}$ must really exist, it also shows that there must exist such functor $F$
which differs from $F_{\min}$ even on the subclass ${\mathcal P}$, that is, one which in line with
\eqref{e} obeys
$F({\mathcal H},\omega,
\varrho_\psi)>\sqrt{\langle\omega\psi,\psi\rangle}$, for some mixed density operator $\omega$ and
unit vector $\psi\in {\mathcal H}$.

Thereby, in view of Theorem \ref{g} and \eqref{e}, to
each mixed  density operator $\omega$ on a Hilbert space ${\mathcal H}$ with
$\dim  {\mathcal H}\geq 2$ there have to exist a unit vector $\psi\in {\mathcal H}$ and a functor
$F$ of that kind. In fact, it is obvious that to each mixed $\omega$ on
${\mathcal H}$ the construction of another
$\omega^\prime$ (on the same Hilbert space, for instance) can be achieved and which satisfies
the condition of the
premise to Corollary \ref{kanal}
in respect of $\omega$. Thus, the above argument by
Corollary \ref{kanal} gets applicable, in either case of mixed $\omega$.
Hence, from the latter and Corollaries \ref{nurJoz}/\ref{kanal} 
we may summarize as follows:
\begin{theorem}\label{nogo}
For each functor $F$ obeying \eqref{ad1}--\eqref{ad6} there exist  
$\{{\mathcal H}^\prime,\omega^\prime,\varrho_{\psi'}^\prime\}, 
\{{\mathcal H},\omega,\varrho_\psi\}\in {\mathcal P}$ such that 
$F({\mathcal H}^\prime,\omega^\prime,\varrho_{\psi'}^\prime)\geq F({\mathcal H},\omega,\varrho_{\psi})$ 
but $\{{\mathcal H}^\prime,\omega^\prime,\varrho_{\psi'}^\prime\}\not\ll_* 
\{{\mathcal H},\omega,\varrho_\psi\}$ holds.
\end{theorem}
After that negative 
result, the second best one can do is to look for  
proper functorial subclasses of ${\mathcal P}$ and containing ${\mathcal P}_0$ 
on which the implication in question with Jozsa's fidelity 
$F=F_{\min}$ might have a chance to be true, at least. 
At that point we will stop. The problem will be clearly posed but will not be solved, in this paper.
 
\medskip
\renewcommand{\labelenumi}{(\arabic{enumi})}
\renewcommand{\theenumi}{\arabic{enumi}}
\noindent{\bf{Problem.}}
If any, find and characterize functorial subclasses ${\mathcal P}^\prime$, ${\mathcal P}_0
\subsetneqq {\mathcal P}^\prime\subsetneqq {\mathcal P}$, such that for
$\{{\mathcal H}^\prime,\omega^\prime,\varrho_\varphi^\prime\},\{{\mathcal H},\omega,\varrho_\psi\}\in
{\mathcal P}^\prime$ the following assertions
were equivalent:
\begin{enumerate}
\item\label{pro1}
$\ \ \ \ \ \{{\mathcal H}^\prime,\omega^\prime,\varrho_\varphi^\prime\}\ll_*\{{\mathcal H},\omega,\varrho_\psi\}$\,;
\item\label{pro3}
$\ \ \,d_{\mathrm{B}}({\mathcal H}^\prime,\omega^\prime,\varrho_\varphi^\prime)\leq
d_{\mathrm{B}}({\mathcal H},\omega,\varrho_\psi)$\,;
\item\label{pro2}
$F_{\min}({\mathcal H}^\prime,\omega^\prime,\varrho_\varphi^\prime)\geq
F_{\min}({\mathcal H},\omega,\varrho_\psi)$\,.
\end{enumerate}
\begin{remark}\label{vorletz}
(1) The category of {\em{cp}}-stochastic linear mappings is one of the generic 
generalizations of the category of stochastic operators/matrices as known from classical 
probability theory. In this sense then the category of functors $F$ obeying 
\eqref{ad1}--\eqref{ad6} is a noncommutative generalization of what is known as 
functionals of type of relative entropy, see \cite{aluh:81.2,albe:88} and  
bibliographies therein. 

(2) Mention that if we restrict the 
dimension of the occuring Hilbert spaces in all objects $\{{\mathcal H},\omega,\varrho\}$ 
by a common finite upper bound, as a special situation the cases with $\varrho=\frac{1}{\dim{{\mathcal H}}}{\mathbf{1}}$ might be considered. The {\em{cp}}-stochastic linear mappings 
acting among those objects then generalize doubly stochastic matrices, and  
in restriction of $\ll_*$ to these special situations with 
$\{{\mathcal H},\omega,\frac{1}{\dim{{\mathcal H}}}{\mathbf{1}}\}$ we arrive at the so-called 
`order structure of states', which is the 
noncommutative version of classical majorization theory, essentially, and which goes back to A.~Uhlmann, 
see \cite{Uh71b,Uh72b,Uh73b}.
 
(3) Also, in view of property \eqref{ad2} (unitary invariance), which corresponds to the classical 
permutational invariance, and by functional calculus, 
in restriction to such situations the system of functors $F$ simply reduces to 
a sufficient system of {\em{Schur-concave functions}}, in the sense of classical majorization 
theory, and Corollary \ref{e0} and Theorem \ref{g} in this special context then 
literally can be concluded also  
from the fundamentals about majorization ordering in classical majorization theory, 
see \cite{AlUh:82} for an overview on unitary mixing and doubly stochastic 
mappings.

(4) On the other hand, and in contrast to the previously discussed types of reductions, 
it does not make any sense to try to find out an equivalent to \eqref{ad1}--\eqref{ad6} 
closed characterization of our functors $F$ and thereby reading in terms of their behavior 
on arguments $\{{\mathcal H},\omega,\varrho\}$, 
with arbitrarily large, but finite dimensions $\dim {\mathcal H}< \infty$, exclusively. 
This is due to the special form of property \eqref{ad6} (structural continuity), 
which makes that reference to infinite 
dimensionality is inherit of that characterization from the very beginning and cannot be circumvented. 

(5) Clearly, this {\em{structural continuity}} is the hardest part to prove, for a given functor. Thus, it 
is important to know functors of that type. One only knows a few series of such functors explicitely, 
among them Jozsa's fidelity, and some other types of relative entropies.
Consequently, and in order to get the doors opened for true applications of $\ll_*$, the hope is 
that questions like the above stated Problem under suitable restrictions on the domain 
(e.g.~low dimensional spaces) would admit proper solutions of practical relevance.   
 
\end{remark}
\subsubsection*{Fidelities in quantum physics--summary}
In a quantum mechanical context, there is no doubt about estimating the accuracy of
transmission through a quantum channel, since these devices are processing within
vector states. In fact, for two quantum (communication) channels
$\gamma:\{{\mathcal H},\varrho_{\rm{in}},\varrho_{\rm{out}}\}$ and $\gamma^\prime:\{{\mathcal H}^\prime,\varrho_{\rm{in}}^\prime,\varrho_{\rm{out}}^\prime\}$ it is naturally to say that the
accuracy of transmission
through $\gamma^\prime$ (at $\varrho_{\rm{in}}^\prime$) is better than the accuracy of transmission
through $\gamma$ (at $\varrho_{\rm{in}}$) if $\varrho_{\rm{in}}^\prime$ and $\varrho_{\rm{out}}^\prime$ are
more neighbouring states than $\varrho_{\rm{in}}$ and $\varrho_{\rm{out}}$,
in the quantum mechanical
sense \cite{Barg:64}, that is, when
$\pmb|${\boldmath{$\psi^\prime$}}-{\boldmath{$\varphi^\prime$}}$\pmb|$ $\leq$ $\pmb|${\boldmath{$\psi$}}-{\boldmath{$\varphi$}}$\pmb|$
holds, for the rays of wave functions $\psi^\prime,\varphi^\prime\in  {\mathcal H}^\prime$ and $\psi,\varphi\in {\mathcal H}$, representing
$\varrho_{\rm{in}}^\prime$, $\varrho_{\rm{out}}^\prime$ and
$\varrho_{\rm{in}}$, $\varrho_{\rm{out}}$, respectively. According to \eqref{i}/\eqref{j},
the latter means $d_{\mathrm{B}}(\varrho_{\rm{in}}^\prime,\varrho_{\rm{out}}^\prime)
\leq d_{\mathrm{B}}(\varrho_{\rm{in}},\varrho_{\rm{out}})$, which
is also the same as $\{{\mathcal H}^\prime,\varrho_{\rm{in}}^\prime,\varrho_{\rm{out}}^\prime\}\ll_*\{{\mathcal H},\varrho_{\rm{in}},\varrho_{\rm{out}}\}$. And equivalently then
$F_{\min}({\mathcal H}^\prime,\varrho_{\rm{in}}^\prime,\varrho_{\rm{out}}^\prime) \geq F_{\min}({\mathcal H},\varrho_{\rm{in}},\varrho_{\rm{out}})$ is fulfilled. Thus, the arising
(pre)ordering of channels in the context at hand can be governed by using a single
`fidelity', which may be chosen as $F=F_{\min}$, for instance.

Clearly, as far as only channels of the class ${\mathcal P}_0$ are concerned, that choice
for `fidelity' might be favoured by mathematical simplicity only. But
there are also substantial hints as to make just that choice for 
`fidelity' and which relate to the functorial class ${\mathcal P}$, 
and which are emphasizing on the exposed r{\^o}le the functor $F_{\min}$ is 
playing in that context.

For instance in quantum chemistry, in application to simple molecular systems
with nondegenerate ground state (let say, with wave function $\psi$),
methods based on (locally) maximizing the
overlap integral $S_\psi(\varphi)=|\int d\/\mu \,\overline{\varphi}\psi|=
|\langle \psi,\varphi\rangle|$ have been widely accepted
and proved successful \cite{Wein:70}. In starting
from an approximate solution $\varphi_0\not={\mathrm{e}}^{{\mathsf{i}}\alpha}\psi$ of the ground
state Schr\"odinger equation,
the idea is to look for trial wave functions $\varphi$ obeying
$S_\psi(\varphi)> S_\psi(\varphi_0)$,
or density functions/density operators $\omega$ with
$S_\psi(\omega)> S_\psi(\varphi_0)$ even, and view them
as better approximations to the ground state than the one given by $\varphi_0$, from the physical
point of view. Thereby,
the extension $S_\psi(\omega)=\sqrt{\langle\omega\psi,\psi\rangle}$ of the overlap
integral is used. The aim then is to pass over from using $\varphi_0$ to a best possible
(maximizing) $\omega$, under boundary conditions
reflecting also those structural informations (symmetries, ground state energy,
gaps in the spectrum etc.) available on the system.
Clearly, since $\psi$ usually is not known explicitly,
one cannot argue with $S_\psi(\omega)$ directly. Usually, instead one then is argueing
with the help of bounds for $S_\psi(\omega)$ and inheriting
from the exact ground state only those informations on the system encoded in the
mentioned boundary conditions \cite{Wein:70,AlHe:89}.
Proceeding `overlapping' this way also quantitatively often is
amazingly successful, and thus `fidelity' according to \eqref{gg} seems to be well-recommended over ${\mathcal P}$, also from this empirical point of view. On the other hand, note that the just discussed 
optimization procedures in quantum chemistry take place under restricted boundary conditions 
within ${\mathcal P}$. These however, exactly correspond to those in the premises of 
Theorem \ref{rein}. Hence, we may take Theorem \ref{rein} 
as a theoretical explanation for diversity and robusteness of the variational methods,
and why overlapping based on $S_\psi(\omega)=\sqrt{\langle\omega\psi,\psi\rangle}$ works so well, 
also quantitatively.  

Relating the whole class ${\mathcal P}$ things are more involved. We now have 
Theorem \ref{nogo}, saying that 
no single functor of the class obeying \eqref{ad1}--\eqref{ad6} can do the job 
of a distinguished fidelity, and since Jozsa's $F_{\min}$ belongs to that class, 
we have a problem, now. At least, this will be so as far as  
notions like `more neighbouring states' and `better accuracy' are suspected 
to be governed by the preordering $\ll_*$, generally (even not only on ${\mathcal P}$). 

It is proposed in this 
paper to take the latter as the working hypothesis, saying that two density operators 
$\omega^\prime$ and $\varrho^\prime$ on a Hilbert space ${\mathcal H}^\prime$
are {\em{more neighbouring states}} than another two density operators 
$\omega$ and $\varrho$ on another Hilbert space ${\mathcal H}$ whenever 
$\{{\mathcal H}^\prime,\omega^\prime,\varrho^\prime\}\ll_* \{{\mathcal H},\omega,\varrho\}$ 
holds. And consequently, on comparing  
transmissions through quantum channels $\gamma^\prime$ (at
$\varrho_{\rm{in}}^\prime$) and $\gamma$ (at
$\varrho_{\rm{in}}$), the transmission through the former will be considered to be of {\em{better 
accuracy}} than the transmission through the latter if $\{{\mathcal
H}^\prime,\varrho_{\rm{in}}^\prime,\varrho_{\rm{out}}^\prime\}\ll_*\{{\mathcal
H},\varrho_{\rm{in}},\varrho_{\rm{out}}\}$ is fulfilled. 

In view of Theorem \ref{g}, under that notion of accuracy all functors obeying 
axioms \eqref{ad1}--\eqref{ad6} provide us with a 
sufficient system of fidelities. We will not compare in detail our proposed system 
of axioms \eqref{ad1}--\eqref{ad6} for fidelities with different other settings of axioms, 
notably the ones posed in \cite{Jozs:94}, in this paper, but instead remark that 
fidelity \`{a} la Uhlmann/Jozsa in both cases arises as the minimal element 
(see \eqref{e} with respect to axioms \eqref{ad1}--\eqref{ad6}).
Clearly, one has to restrict $\ll_*$ to domains where the concepts can be 
reasonably quantified, that is, literally speaking, where the full information on that 
matter can be gained by monitoring a `few' standing (and thus distinguished) fidelities, exclusively. 

There are some indications saying that for this sake and for testing the hypothesis the class ${\mathcal P}$ 
might be a good choice as a starting domain. Namely, 
in order to deal with $\ll_*$ on ${\mathcal P}$ (or in order to study  
the bounds a solution ${\mathcal P}^\prime$ might be subject to), the likely more promising strategy 
than that working with all the functors $F$ could be to try to prove/disprove to given channels in 
${\mathcal P}$ existence of 
a {\em{cp}}-stochastic map $\Phi$ achieving $\varrho_{\rm{in}}^\prime=\Phi(\varrho_{\rm{in}})$ and 
$\varrho_{\rm{out}}^\prime=\Phi(\varrho_{\rm{out}})$, explicitely. 
Thereby, the simplest compromise seems to look for 
solutions of the above stated Problem, that is, to look for proper functorial subclasses  
${\mathcal P}^\prime$ of ${\mathcal P}$, with
${\mathcal P}_0\subsetneqq {\mathcal P}^\prime$, and over which $F_{\min}$ will play 
its part of the distinguised fidelity, again. 

This is a hopeful procedure for several reasons: firstly, we have to our disposal 
those canonical parametrizations within manifolds of {\em cp}-stochastic linear
mappings \cite{krau:71,evle:77},
which naturally come along with the notion of complete positivity \cite{stine:55}.
The second point could be, that the previous together with the fact that on ${\mathcal P}$ both 
input states have to be pure ones, in form of $\varrho_{\rm{in}}^\prime=\Phi(\varrho_{\rm{in}})$ 
are subject to a very restricting condition. 
And as a third point, we can make also use of variational bounds, 
which are known to hold for $F_{\min}$, see \cite[{\sc Corollary 2}\,(1)--(5)]{aluh:00.1}. 
In concert, these then might allow for estimating boundaries for $\varrho_{\rm{out}}^\prime$ 
relative to $\varrho_{\rm{out}}$, at least.
For low dimensions or commuting density matrices, such procedures might work well, 
see analogous procedures in \cite{AlUh:80,Zylk:81} and arguments raised in the proof of Corollary \ref{kanal}. 
\end{appendix}

\end{document}